\documentclass[12pt]{article}
\usepackage{mathtools}
\usepackage{amsmath, amsthm, amssymb,multirow, amsfonts, amsopn, bm, bbm}
\usepackage{relsize} % make large math symbols
\usepackage{hyperref,tikz,epsfig}
\usepackage[edges]{forest}
\usepackage{tikz}
\usetikzlibrary{trees}
\usepackage{mwe}
\usepackage{subfig}
\usepackage{breqn} % break long math into lines
\usepackage{forest}
\usepackage{graphicx} % Required for inserting images
\usepackage[margin=2.5cm]{geometry}
\usepackage{setspace}
\usepackage{url}
\usepackage{authblk}
\usepackage{subcaption}
\title{\large A Real Data-Driven, Robust Survival Analysis on Patients who Underwent Deep Brain Stimulation for Parkinson's Disease by Utilizing Parametric, Non-Parametric, and Semi-Parametric Approaches}
\author[1*]{\small Malinda Iluppangama}
\author[2]{\small Dilmi Abeywardana}
\author[2]{\small Chris P. Tsokos}
\affil[1]{\small Department of Mathematics and Statistics, Loyola University, Baltimore, MD, USA.}
\affil[2]{\small Department of Mathematics and Statistics, University of South Florida, Tampa, FL, USA.}
\affil[*]{\small Corresponding author: Malinda Iluppangama (miluppangama@loyola.edu)}
\date{}
\begin{document}
\maketitle
\vspace{-1cm}
\begin{abstract}

Parkinson's Disease (PD) is a devastating neurodegenerative disorder that affects millions of people around the globe. Many researchers are continuously working to understand PD and develop treatments to improve the condition of PD patients, which affects their day-to-day lives. In recent decades, the treatment, Deep Brain Stimulation (DBS), has given promising results for motor symptoms by improving the quality of daily living of PD patients. In the methodology of the present study, we have utilized sophisticated statistical approaches such as Nonparametric, Semi-parametric, and robust Parametric survival analysis to extract useful and important information about the long-term survival outcomes of the patients who underwent DBS for PD. Finally, we were able to conclude that the probabilistic behavior of the survival time of female patients is statistically different from that of male patients. Furthermore, we have identified that the probabilistic behavior of the survival times of Female patients is characterized by the 3-parameter Lognormal distribution, while that of Male patients is characterized by the 3-parameter Weibull distribution. More importantly, we have found that the Female patients have higher survival compared to the Male patients after conducting a robust parametric survival analysis. Using the semi-parametric COX-PH, we found that the initial implant of the right side leads to a high frequency of events occurring for the female patients with a bad prognostic factor, while for the male patients, a low events occurs with a good prognostic factor. Furthermore, we have found an interaction term between the number of revisions and the initial size of the implant, which increases the frequency of events occurring for the Male patients with a bad prognostic factor.\newline

\end{abstract}

\section{Introduction}

 Parkinson's Disease (PD) is a devastating neurodegenerative disorder that affects the brain neuron system, and medical professionals and research scientists have been studying PD for a long period of time. However, the cause of PD still remains undetected. If the cause of PD is unknown, it is called idiopathic PD and is the most common type of PD. Also, patients who get PD before 50 years of age are known as having young-onset Parkinson's disease. Unintended/Uncontrollable movements such as shaking, stiffness, and loss of balance are some of the major symptoms of PD, among others. Symptoms of PD start slowly, progress over time, and finally get worse over time. Patients in the later stage of PD are unable to walk or talk,~\cite{PD_cause},~\cite{Rmayya_dbs}. Moreover, PD is a well-known symptom of the devastating dementia disease. 
 
 Over 10 million people around the globe suffer from PD, and more than 50,000 people are diagnosed with PD each year, only in the United States. However, these numbers do not represent thousands of undetected cases of PD around the world,~\cite{PD_challange}. In our previous studies, we have utilized a real data-driven machine learning modeling approach to detect PD patients with a very high degree of accuracy,~\cite{My_Machine_learning_PD}, and invented a real data-driven analytical model to identify the causes of PD through Unified Parkinson's Disease Rating Scale (UPDRS),~\cite{tsokos2025statistical_enclyclopedia}. 

 Research scientists have been working over the last few decades on therapies and drugs to slow down or cure this devastating disease. Lepadopa and Dopamine agonist are the most common medications among others to slow down this devastating disease,~\cite{Levodopa_PD_medication}. Also, Deep Brain Stimulation (DBS) is a surgical therapy that has been used as a treatment to improve the movement functions and the quality of life of patients with movement disorders. Since movement disorders are one of the major symptoms of PD, DBS is widely used to treat PD. DBS is one of the most powerful innovative treatments that has shown promising results for the movement symptoms of advanced PD, such as bradykinesia, essential tremor, and other motor complications of PD. Also, it reduces other medication usages of an individual patient with PD,~\cite{parkinsonsfoundation},~\cite{PD_dbs_imporve},~\cite{Günther_Deuschl_et_al_DBS_improvement}. 

In the literature, we found that several studies have confirmed that the survival times of PD patients increase with DBS therapy compared to the PD patients who are taking other treatments,~\cite{Ngoga_DBS_survival}. For instance, we found several articles that have been trying to analyze the long-term survival outcome of PD patients who underwent DBS, and all the articles uniformly conclude the significant overall improvement of the patients' motor symptoms. Also, medication like levodopa equivalent daily dose was reduced over 55\% for 5 years of follow-up after DBS,~\cite{Fasano_8_year_dbs_followup},~\cite{Schüpbach_5_year_folloup_DBS}. Furthermore, we found that a greater number of publications show an over 98\% survival rate after 5 years of DBS treatment, and also show a very high survival rate even after 3 years of DBS treatment,~\cite{Kim_mortality_DBS},~\cite{Ryu_DBS_mortality},~\cite{Kim_A_DBS_mortality}. In another study, the estimated long-term (10 years) survival probability was identified as 51\%,~\cite{Rmayya_dbs}. However, in the literature, most researchers utilized the most widely used nonparametric Kaplan-Meier approach to perform the analysis and made conclusions about the survival of the patients who underwent DBS for PD, but it is neither a powerful nor a robust method. Furthermore, researchers utilized the semiparametric Cox-Proportional Hazards approach without addressing its underlying important assumptions and made conclusions about the survival of the patients, and these results may be misleading because of the inappropriateness/invalidness of the developed models with unsatisfied underlying assumptions.

In this study, we analyze the survival of the patients who underwent DBS for PD more sophistically and investigate the sustainability of the long-term survival of PD by utilizing well-defined, robust statistical methodologies such as Nonparametric, Semiparametric, and Parametric. Thus, the rest of this paper is organized as follows. Sections 2 and 3 consist of a descriptive analysis of the data and a brief introduction to sophisticated statistical methodologies, respectively. Finally, Sections 4 and 5 discuss the results and the conclusions of the study, highlighting the usefulness of this study and concluding remarks.

%\newpage
\section{The Real Data and Descriptive Analysis}

%\subsection{Data}
The real data consists of information on individuals who underwent the DBS treatment for PD starting from 1999 to 2006, with at least 10 years of follow-up at Pennsylvania Hospital. We had the opportunity to get access to this data through the contact of Dr.Ashwin Ramayya (MD),~\cite{Rmayya_dbs}. The data consists of a total of 320 individuals who underwent DBS for PD. However, we considered only complete observations (deceased patients) for our study of survival analysis, and it consists of only 149 individual patients after removing the censored observations and the missing observations without using any data imputation method to replace those missing survival times of the individual patients. 

\begin{figure}[h]  
\begin{center}
\includegraphics[height=9cm]{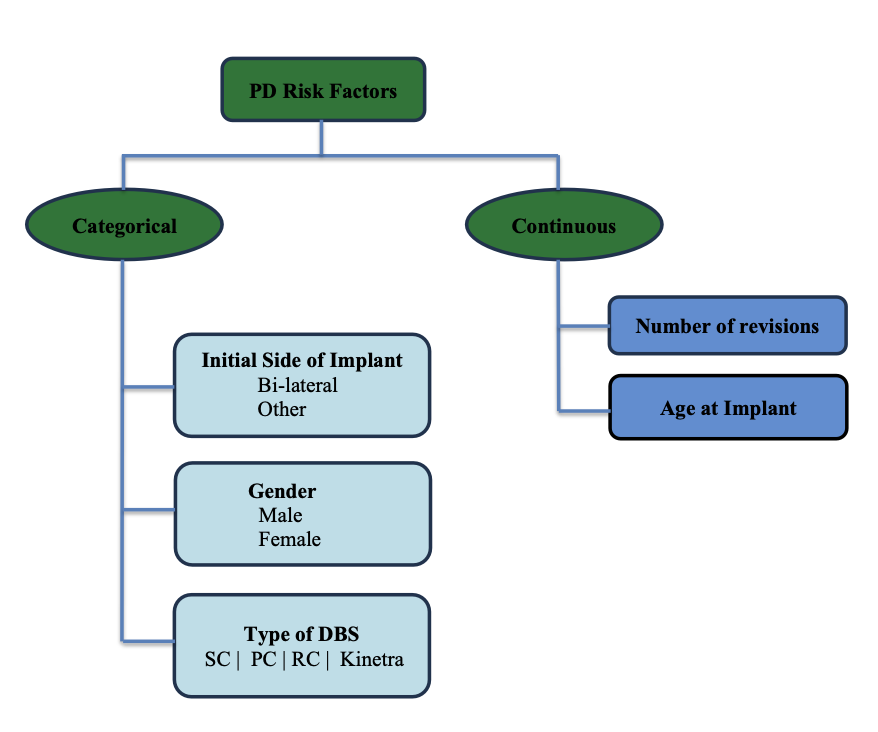}
\end{center}
\caption{Data Diagram of Risk Factors}
\label{Figure 3}
\end{figure}

Age, Age at Implant, Gender, Survival time (in months), Initial implant side, and Number of revisions are the six individual covariates that we have used to develop the present study. The final data frame consists of individual patients in the range of age from 40 years to 93 years old, and the mean age of the patients who underwent DBS surgery is 64.41 years ($\pm$ 8.59 standard deviation). Also, 73\% of the total individual patients are Males. Furthermore, 69\% of the patients underwent bilateral DBS for PD. The following data diagram, Figure~\ref{Figure 3}, and Table~\ref{table1} illustrate the exploratory data analysis for some continuous covariates.

%\begin{figure}[h]  
%\begin{center}
%\includegraphics[height=5cm]{Figures/DBS.png}
%\end{center}
%\caption{Deep Brain Stimulation}
%\label{Figure 2}
%\end{figure}

\begin{table}[!h]
\begin{center}
\begin{tabular}{ccccc}
    \hline   
      Risk Factor & Minimum  & Maximum & Mean & Standard Deviation \\
     \hline
     \hline
     Age at Implant & $33$ & $86 $ & $64.41$ & $8.59$ \\
          % we can use this when the table elements are too long
     
     Survival time & $3$ & $212$ & $98.80$ & $42.12$ \\
     \hline
\end{tabular}
\end{center}
\caption{Descriptive Statistics of Continuous Risk Factors}
\label{table1}
\end{table}

\section{Methodology}
In this study, we utilize the well-known non-parametric  Kaplan-Meier (KM) approach, robust Parametric approach, and widely used semi-parametric multivariate Cox-PH method to analyze the long-term survival outcomes of the individuals who underwent DBS for PD. In the following section, each of the proposed models and their usefulness are briefly discussed.

\subsection{Kaplan-Meier Survival Analysis}

Kaplan-Meier (KM) estimation of the survival time is a fully non-parametric approach for analyzing the time of occurrence of the event of interest (time until death), and it was first introduced in 1958,~\cite{KM}. KM is a well-known approach in the field of survival analysis and reliability analysis. However, KM is not as powerful nor robust as the Parametric or Bayesian survival analysis since it does not make any distributional assumption about the underlying survival time. KM involves estimating the survival probabilities of patients at a certain point in time as follows,

\begin{equation*}
    S_t = \frac{Number \, of \,patients\, living \,at \,the\, start \,(t_0) - Number\, of\, patients\, dies \,at \,time\, t}{Number\, of\, patients \,living\, at\, the\, start\, (t_0)}.
\end{equation*}

Suppose that there are $n$ individual patients at the start of time ($t_0$) and their respective observed event times (died at time $t_i$) are $t_1 < t_2 < t_3 <...<t_k$ with $d_i$ units failing at time $t_i$. Then the KM estimator of the survival function $S_{KM}(t)$ is given by following equation,

\begin{equation}
    S_{KM}(t) = \prod_{i:t_i<t} \frac{(n_i - d_i)}{n_i} = \prod_{i:t_i<t}(1-\frac{d_i}{n_i})
    \label{Eq_KM},
\end{equation}

where $t>0$,
$n_i$ is the number of patients at risk at time $t_i$ of the study,
$d_i$ is the number of patients who died at time $t_i$. We proceed and calculate survival probabilities using KM estimators given by the equation~\ref{Eq_KM} under the following assumptions,~\cite{KM_Basics}, ~\cite{KM_Explain},

\begin{itemize}
    \item Patients who have been censored have the same survival prospects as those who continue the follow-up.
    \item Survival probabilities are the same for individuals who were recruited early and late in the study.
    \item Event happens at the time specified.
\end{itemize}

\subsection{Parametric Survival Analysis}

In this section, we briefly discuss the more powerful and robust parametric survival analysis and its usefulness. 

Let T be a non-negative random variable, which measures the survival time of an individual patient, and its probabilistic behavior is characterized by the well-defined probability density function $f(t)$. Then the probability that an event occurs prior to time t can be defined as the cumulative distribution function and is given below by Equation~\ref{Cumulative_1},

\begin{equation}
    F(t) = P(T \leq t) = \int_{0}^{t} f(t)\,dt\quad, \quad\quad t>0,
    \label{Cumulative_1}
\end{equation}

where $f(t)$ is the probability density function of the survival time. 

The survival of an individual patient after time t is given by the Survival Function, and it can be derived from the above CDF. The generalized analytical form of the survival function can be derived as follows in Equation~\ref{survival_function_1}.

\begin{equation}
    S(t) = 1 - F(t) = P(T > t) = \int_{t}^{\infty} f(t)\,dt\quad, \quad\quad t>0.
    \label{survival_function_1}
\end{equation}

One can utilize Equation~\ref{Cumulative_1} and Equation~\ref{survival_function_1} to derive the hazard function to measure the chances of instantaneous death of a given patient at any time t as follows in Equation~\ref{hazard_fun_def_1},

\begin{equation}
    h(t) = \lim_{\delta t\to\ 0} \frac{P(t < T \leq t +\delta t | T > t)}{\delta t} = \frac{f(t)}{S(t)}.
    \label{hazard_fun_def_1}
\end{equation}

\subsection{Multivariate Cox-Proportional Hazards Model}

In the context of survival/reliability analysis, the hazard function plays a major role. It explains the instantaneous risk of an event at time $t$. That is, the probability of the desired event occurring any time after time $t$ is the hazard function under the assumption that the individual observation has not experienced the desired event until time $t$.\\

Let $T$ be a random variable that represents the survival time of a given patient. Then the probability that an event occurs prior to or at time $t$ is defined by the cumulative distribution function (CDF) and is given below,

\begin{equation}
    F(t) = P(T \leq t) = \int_{0}^{t} f(t)\,dt\quad, \quad\quad t>0,
    \label{Cumulative}
\end{equation}
 where the probability density function can be derived as follows,
\begin{equation}
    f(t) = \frac{dF(t)}{dt}.
    \label{probability_function}
\end{equation}
The probability of a given patient surviving after time $t$ (Survival function) can be defined by the following Equation~\ref{survival_function}, 

\begin{equation}
    S(t) = 1 - F(t) = P(T > t) = \int_{t}^{\infty} f(t)\,dt\quad, \quad\quad t>0.
    \label{survival_function}
\end{equation}

Utilizing Equation~\ref{probability_function} and Equation~\ref{survival_function}, the hazard function for an individual patient at time $t$ can be derived as follows,

\begin{equation}
    h(t) = \lim_{\delta t\to\ 0} \frac{P(t < T \leq t +\delta t | T > t)}{\delta t} = \frac{f(t)}{S(t)}.
    \label{hazard_fun_def}
\end{equation}

To estimate the hazard function, the Cox Proportional Hazards Model (Cox-PH) was introduced by D. R Cox in 1972,~\cite{COX}, which specifies the hazard function directly rather than deriving it as the ratio of the probability density function and the survival function. The analytical form of the Cox-PH model is given by Equation~\ref{coxph_model},

\begin{equation}
\label{coxph_model}
    h_i (t) = h_0 (t) exp\left(\sum_{j=1}^{p} \beta_j X_{j} + \sum_{j=1}^{p} \alpha_{jK} X_{j}*X_{K} \right ),
\end{equation}

where $h_0 (t)$ is the baseline hazard function, and it depends on only the survival time. $\beta_j$ is the coefficient parameter for $j^{th}$ covariate and  $\alpha_{jk}$ is the coefficient parameter for the interaction between $j^{th}$ and $k^{th}$ covariates. In the Cox-PH model, the baseline hazard function $h_0 (t)$ is unspecified and it is treated as a non-parametric function of time $t$. However, the Cox-PH model assumes the time independence of the covariates, linearity in the covariates, additivity, and proportional hazards. Thus, the Cox-PH model is considered as a semi-parametric survival model,~\cite{cox_dilmi}. To estimate the parameters of the Cox-PH model, we cannot use the full likelihood approach since the baseline hazard function is not specified. Thus, the partial likelihood approach proposed by Cox in 1979 is utilized to estimate the model coefficients of the Cox-PH model,~\cite{Cox_theory}. 

In many situations, researchers are interested in the factor exp(coef) of the model rather than estimating the whole $h_i(t)$, which determines how an individual patient varies as a proportion of the baseline hazard function. Equation~\ref{coxph_model} can be rearranged to find the hazard ratio for a given individual patient as given below,

\begin{equation}
   HR = \frac{h_i(t)}{h_0(t)}  = exp\left( \sum_{j=1}^{p} \beta_j X_{ij}\right).
    \label{hazard_ratio}
\end{equation}

Suppose that we have two types of treatment groups, namely the treatment group (T) and the placebo group (P). Then we can extract the following important and useful information about the two treatment groups using the HR,

\begin{itemize}
    \item HR $> 1 (\beta_i>0$). The treatment group has a higher hazard than the placebo group. (i.e., the placebo group is favored.)
    \item HR $\approx1$. The treatment group and placebo group have no significant difference.
    \item HR $< 1 (\beta_i<0$). The treatment group has a lower hazard than the placebo group. (i.e., the treatment group is favored.)
\end{itemize}

%\newpage
\section{Results}
\label{Results}

Before modeling survival times, we assess whether the median survival time of the Male patients who underwent DBS for PD is statistically significantly different from that of female patients by utilizing hypothesis testing such as the Wilcoxon rank sum test and log-rank test,~\cite{log_rank}, ~\cite{Wilcox}, ~\cite{Wilcox_rank_sum}. In the following section, we have defined our primary null hypothesis and alternative hypothesis.

\begin{center}
$H_0$: Median Survival Time of Male patients is not different from that of Female patients.\\
    Vs
    
$H_1$: Median Survival Time of Male patients is different from that of Female patients.
\end{center}

\begin{figure}[hbt!]
    \centering
    \includegraphics[width=4.5in]{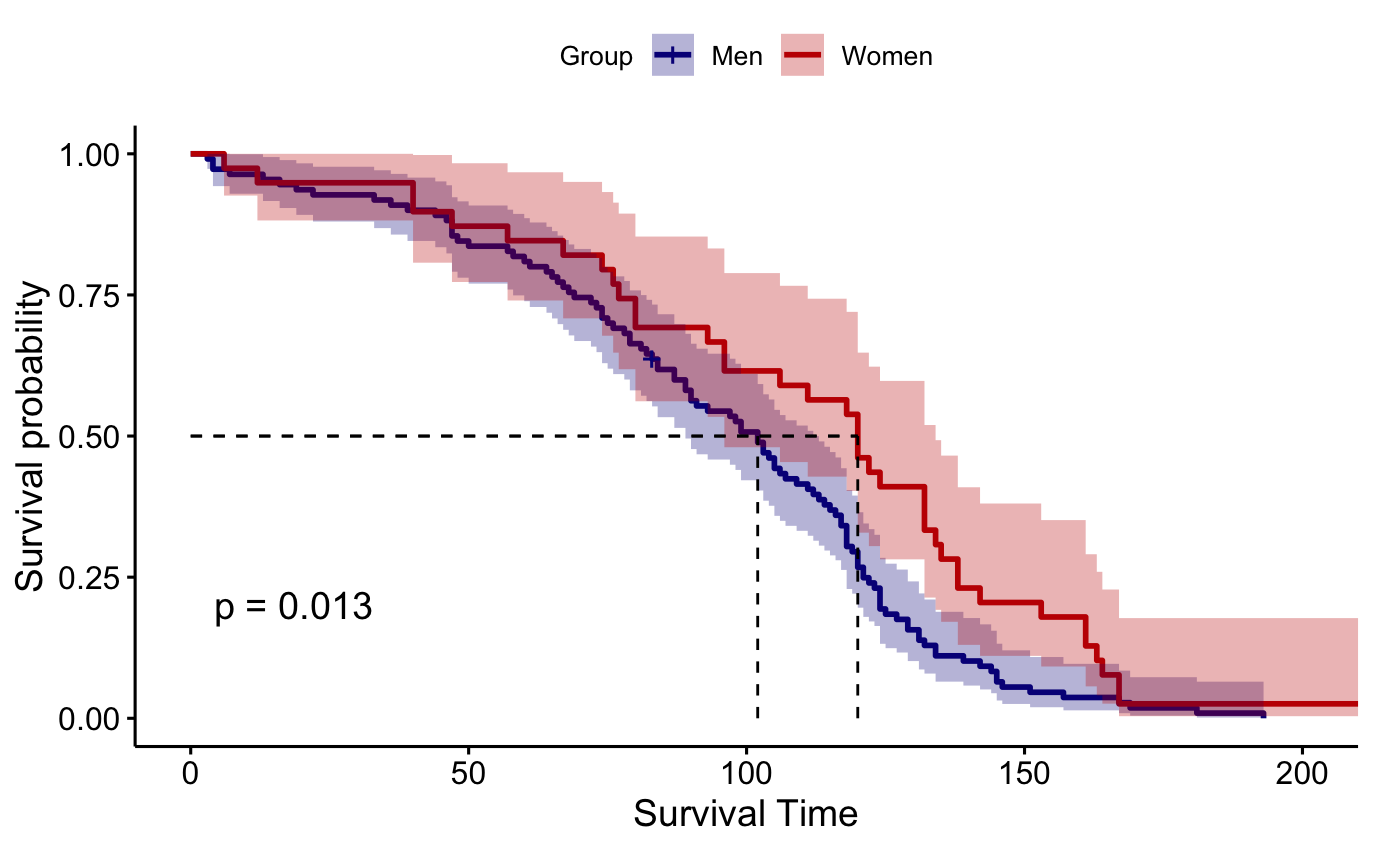}
    \caption{Visualization of Log-Rank Test Results}
    \label{fig:Influence_M1}
\end{figure}

The Wilcoxon rank sum test gave a p-value of 0.029, and for the log-rank test, the p-value is 0.013 as visualized in Figure~\ref{fig:Influence_M1}. Both Wilcoxon and log-rank tests uniformly confirmed that p-values are less than the 5\% significance level. Thus, we could reject the null hypothesis and conclude that the median survival time of male patients is different from that of females. Since the median survival time of males is significantly different from that of females, we consider males and females as two different entities and continue our analysis separately for male patients and female patients.

%\newpage
\subsection{Univariate Survival Analysis}

In this subsection, we summarize the results of parametric and nonparametric univariate survival analysis for PD patients who had DBS.

\subsubsection{Kaplan Meier (KM)}
\begin{figure}[hbt!]
    \centering
    \includegraphics[width=5in]{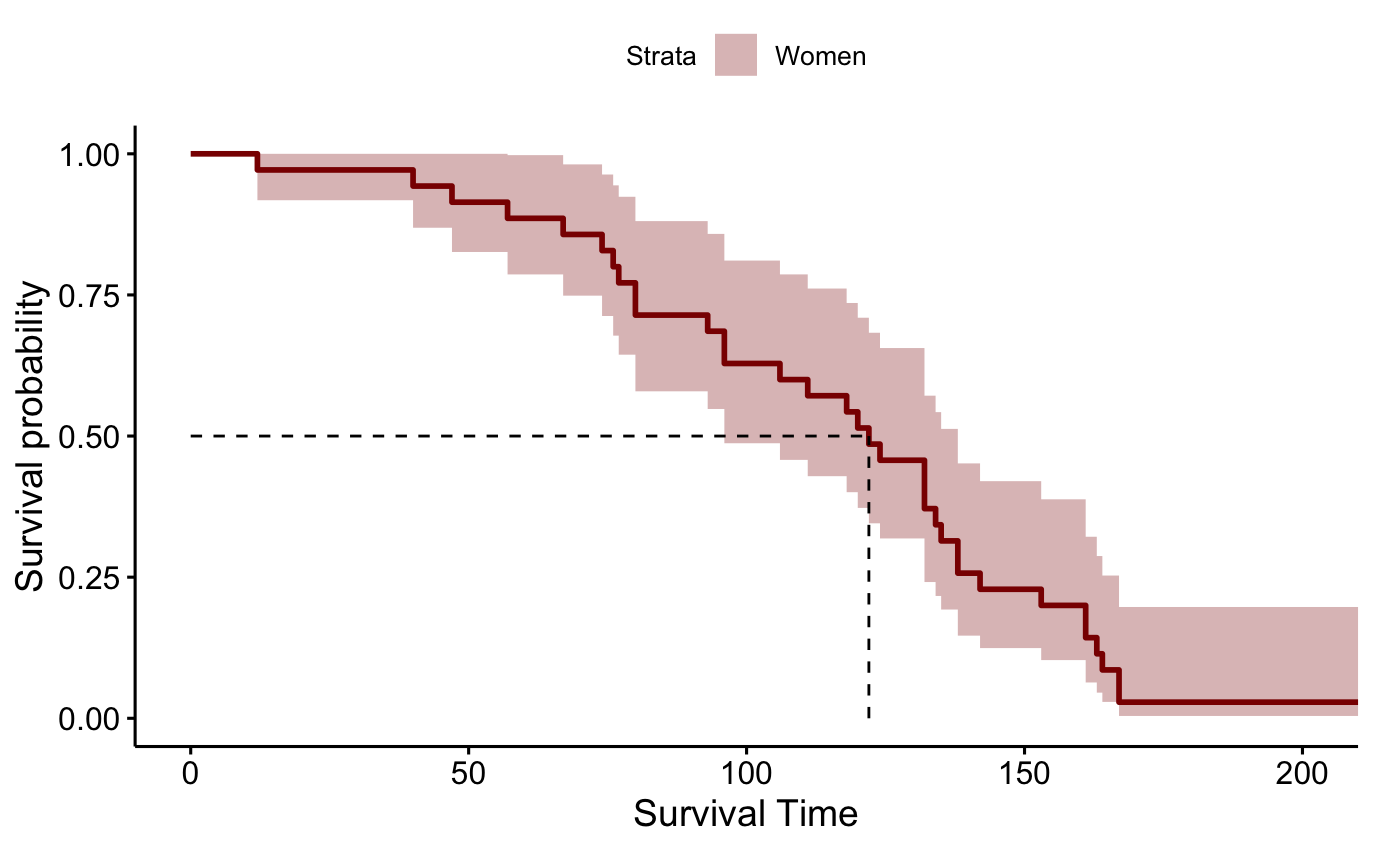}
    \caption{KM Survival Plot for Female Patients.}
    \label{fig:KM_F}
\end{figure}

\begin{figure}[hbt!]
    \centering
    \includegraphics[width=5in]{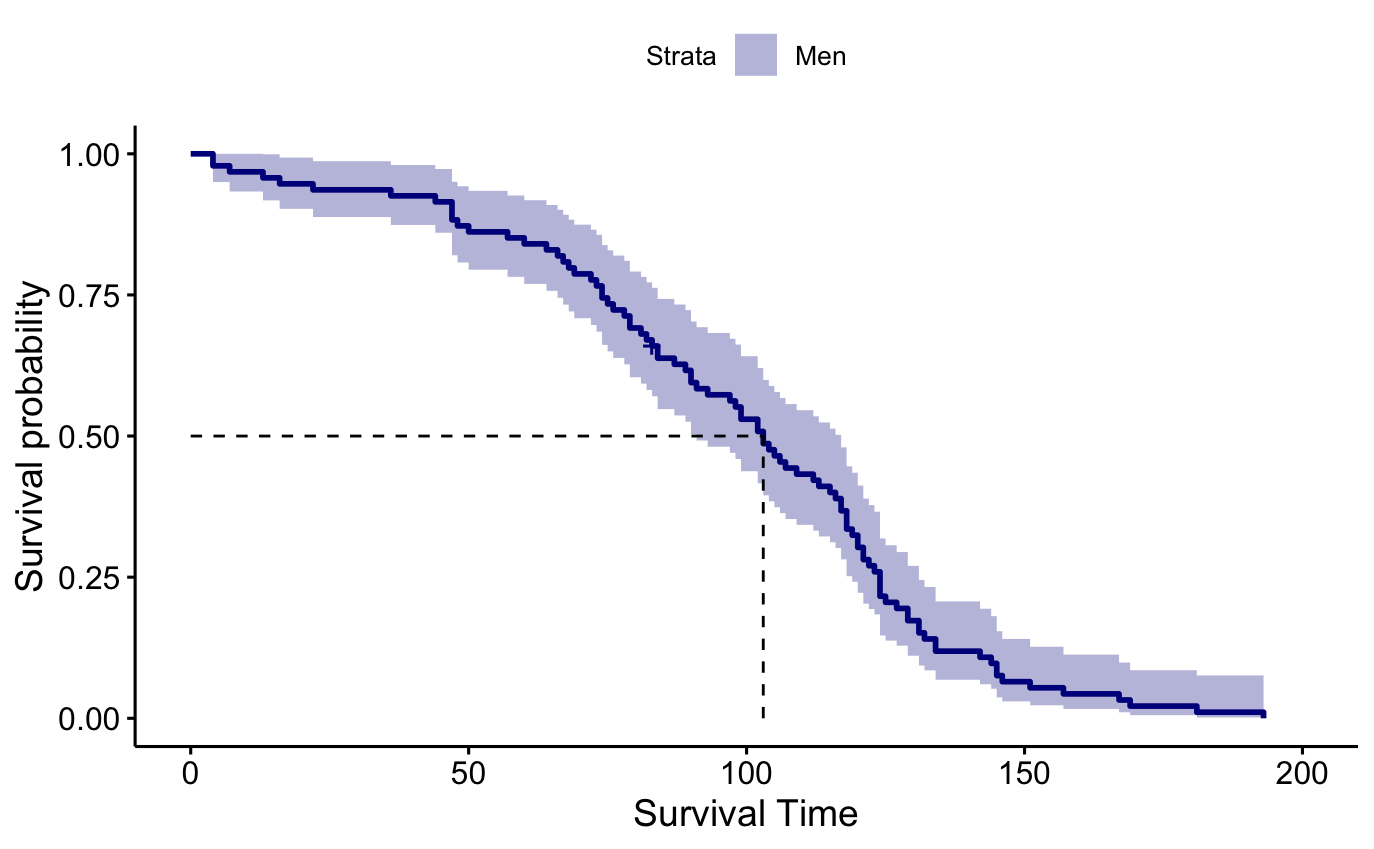}
    \caption{KM Survival Plot for Male Patients.}
    \label{fig:KM_M}
\end{figure}

KM survival probabilities have been estimated for both male patients and female patients separately because initially, we identified that their survival times are statistically significantly different. The KM estimation results for the Male patients and Female patients are summarized in Table~\ref{Kaplan_Meier_Coefficient_ table}  and graphically illustrated by Figure~\ref{fig:KM_F} and Figure~\ref{fig:KM_M} for Females and Males, respectively. 

\begin{table}[!h]
\begin{center}
\begin{tabular}{ccccccc}
    \hline   
      Time & n.risk & n.event & Survival(pr) & std.err & 95\% Lower & 95\% Upper  \\
     \hline
     \hline
     &  &  & \textbf{Male Patients}\\
     \hline
     3 & 109 & 1 & 0.99083 & 0.00913 & 0.97309 & 1.0000 \\
     4 & 108 & 2 & 0.97248 & 0.01567 & 0.94224 & 1.0000\\
     7 & 106 & 1 & 0.96330 & 0.01801 & 0.92864 & 0.9993\\
     13 & 105 & 1 & 0.95413 & 0.02004 & 0.91565 & 0.9942 \\
     16 & 104 & 1 & 0.94495 & 0.02185 & 0.90309 & 0.9888\\
     19 & 103 & 1 & 0.93578 & 0.02348 & 0.89087 & 0.9830\\
     22 & 102 & 1 & 0.92661 & 0.02498 & 0.87892 & 0.9769\\
     33 & 101 & 1 & 0.91743 & 0.02636 & 0.86719 & 0.9706\\
     36 & 100 & 1 & 0.90826 & 0.02765 & 0.85565 & 0.9641\\
     39 & 99 & 1 & 0.89908 & 0.02885 & 0.84428 & 0.9574\\
     44 & 98 & 1 & 0.88991 & 0.02998 & 0.83305 & 0.9507 \\
     46 & 97 & 1 & 0.88073 & 0.03104 & 0.82194 & 0.9437\\
     47 & 96 & 3 & 0.85321 & 0.03390 & 0.78929 & 0.9223\\
     48 & 93 & 1 & 0.84404 & 0.03475 & 0.77860 & 0.9150\\
     50 & 92 & 1 & 0.83486 & 0.03556 & 0.76799 & 0.9076\\
     \hline
     &  &  & \textbf{Female Patients}\\
     \hline
     6 & 39 & 1 & 0.9744 & 0.0253 & 0.9260 & 1.000 \\
     12 & 38 & 1 & 0.9487 & 0.0353 & 0.8820 & 1.000\\
     40 & 37 & 2 & 0.8974 & 0.0486 & 0.8071 & 0.998\\
     47 & 35 & 1 & 0.8718 & 0.0535 & 0.7729 & 0.983 \\
     57 & 34 & 1 & 0.8462 & 0.0578 & 0.7402 & 0.967\\
     67 & 33 & 1 & 0.8205 & 0.0615 & 0.7085 & 0.950 \\
     74 & 32 & 1 & 0.7949 & 0.0647 & 0.6777 & 0.932\\
     76 & 31 & 2 & 0.7692 & 0.0675 & 0.6477 & 0.914\\
     77 & 30 & 1 & 0.7436 & 0.0699 & 0.6184 & 0.894 \\
     80 & 29 & 2 & 0.6923 & 0.0739 & 0.5616 & 0.853\\
     93 & 27 & 1 & 0.6667 & 0.0755 & 0.5340 & 0.832\\
     96 & 26 & 2 & 0.6154 & 0.0779 & 0.4802 & 0.789\\
     106 & 24 & 1 & 0.5897 & 0.0788 & 0.4539 & 0.766 \\
     111 & 23 & 1 & 0.5641 & 0.0794 & 0.4281 & 0.743\\
     118 & 22 & 1 & 0.5385 & 0.0798 & 0.4027 & 0.720\\
     \hline
\end{tabular}
\end{center}
\caption{Kaplan Meier Survival Estimations}
\label{Kaplan_Meier_Coefficient_ table}
\end{table}

When carefully investigating the above KM plots for the two genders, it is clear that the median survival time of female patients is higher than the median survival time of Male patients after performing DBS for PD.

\newpage
\subsubsection{Parametric Survival Analysis}

In the case of Parametric survival analysis, it is essential to identify the probability distribution functions that characterize the probabilistic behavior of the survival time of the Male patients and Female patients separately, since their median survival times are statistically significantly different. Thus, we identified the best-fitted probability distribution functions for both male and female patients that characterize the probabilistic behavior of their survival times and utilized well-defined goodness of fit tests, Kolmogorov-Smirnov, Anderson-Darling, and Chi-Squared to validate the resulting probability distribution functions. The goodness of fit test results are summarized in Table~\ref{Goodness_of_fit_Coefficient_ table}, given below.

\begin{table}[!h]
\begin{center}
\begin{tabular}{cccccc}
\hline
& & & & \textbf{\footnotesize Goodness of Fit Test(p-vales)}\\ 
     \footnotesize Gender & \footnotesize Distribution & Obs & \footnotesize Kolmogorov-Sminorv & \footnotesize Anderson-Darling & \footnotesize Chi-Square  \\
      \hline
     \hline
     \textbf{\footnotesize Male} & \textbf{\footnotesize Weibull (3P)} & \footnotesize 109 & \footnotesize 0.840 & \footnotesize 0.475 & \footnotesize 0.605 \\
     \textbf{\footnotesize Female} & \textbf{\footnotesize Log-normal(3P)} & \footnotesize 39 & \footnotesize 0.429 & \footnotesize 0.527 & \footnotesize 0.924\\
     \hline
\end{tabular}
\end{center}
\caption{Goodness of Fit Test Summary}
\label{Goodness_of_fit_Coefficient_ table}
\end{table}

The goodness of fit test results given by Table~\ref{Goodness_of_fit_Coefficient_ table} uniformly confirmed that the probabilistic behavior of the survival times of the Male patients are characterized by the 3-parameter Weibull distribution and its analytical form is given by the mathematical function~\ref{3p-weibull}, given below. Furthermore, parameters of the Weibull distribution were estimated by using the most commonly applied Maximum Likelihood Estimation (MLE) method,~\cite{Weibull_parameter}. 

\begin{equation}
    f_M(x|\theta,\eta,\tau) =\frac{\eta}{\theta}(x - \tau)^{\eta -1}exp\{-\frac{1}{\theta}(x - \tau)^\eta\} , \hspace{2em} \eta > 0,\, \theta > 0,\, x > \tau,
    \label{3p-weibull}
\end{equation}
where $\theta$, $\tau$, and $\eta$ are Shape, Scale, and Location parameters, respectively. MLE estimates for the model parameters are $\theta$= 4.9593, $\eta$ = 186.21, $\tau$= -76.262. 

Once the parameters are estimated, the shape of the Weibull distribution that characterizes the probabilistic behavior of the survival times of the male patients is graphically illustrated by~\ref{fig:Webill_distribution}, given below.
\begin{figure}[ht!]
    \centering
    \includegraphics[width=4.25in]{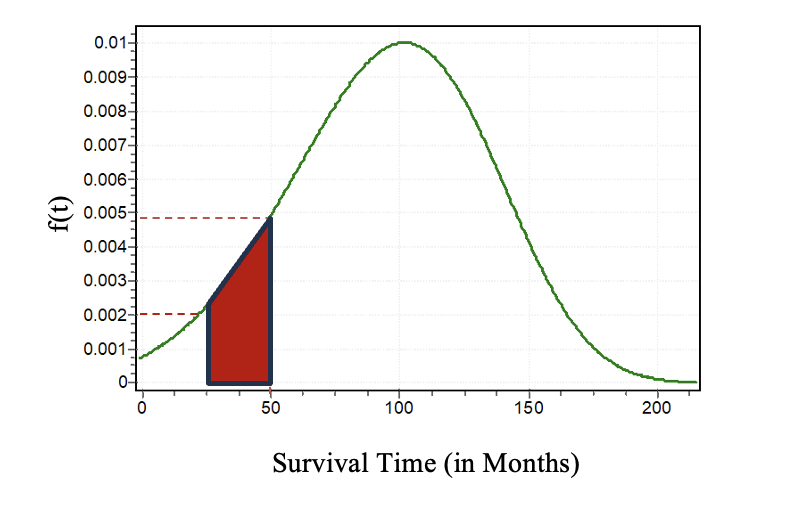}
    \caption{Probability Distribution Function of Survival Times of Male Patients}
    \label{fig:Webill_distribution}
\end{figure}

From Figure~\ref{fig:Webill_distribution}, one can compute the probabilities of the survival times of the patients who underwent DBS for PD between time $t_i$ and $t_{i+1}$. As an illustration, we can compute the probability of survival between 25 months to 50 months for a PD patient who underwent DBS as $P( 25 \leq t \leq 50)$ $\approx 0.00475-0.002 \approx 0.00275$. In Figure~\ref{fig:Webill_distribution}, we can clearly see that the probability of a patient who will survive between 25 months to 50 months is approximately $0.3\%$.

\begin{figure}[hbt!]
    \centering
    \includegraphics[width=5in]{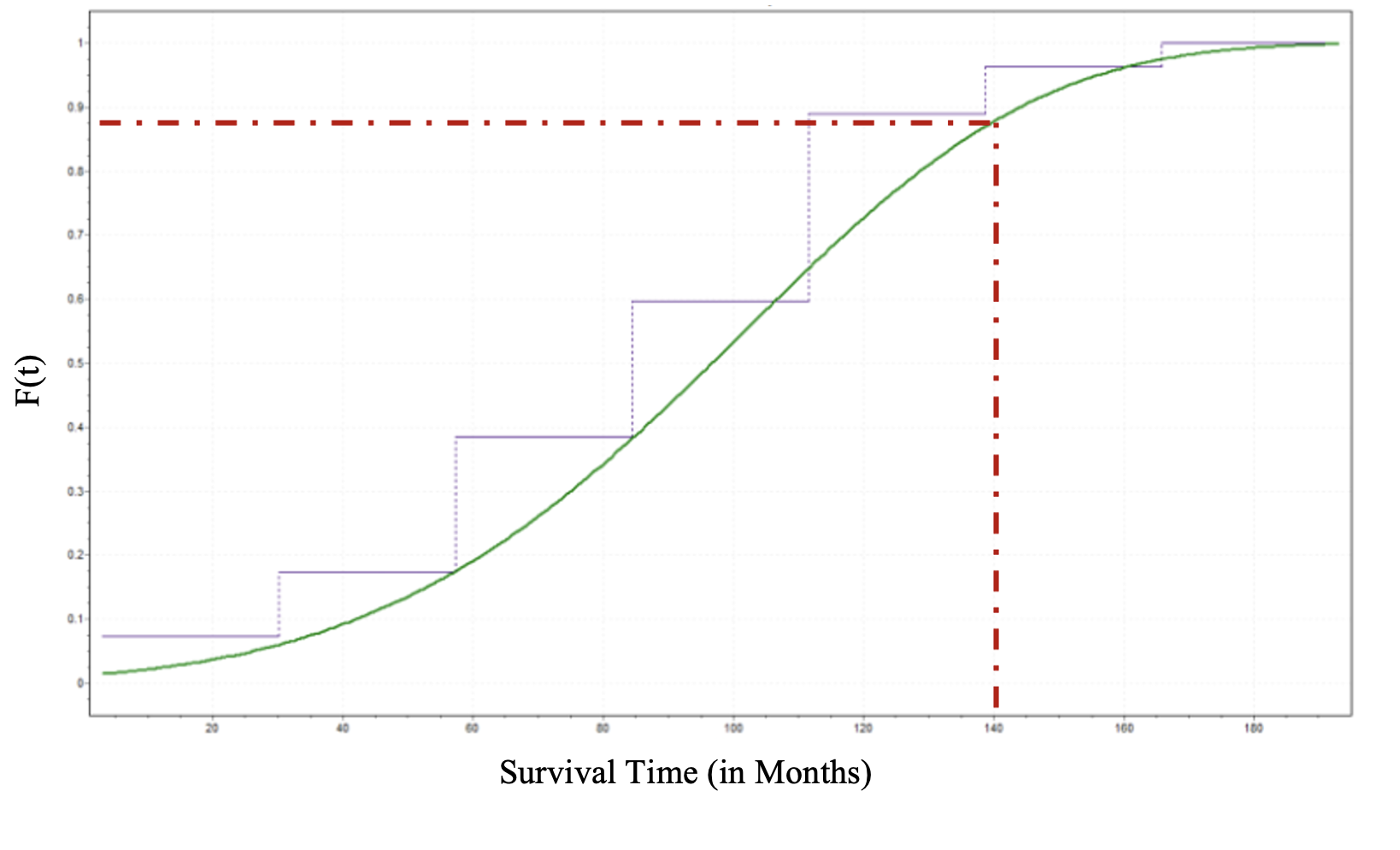}
    \caption{Cumulative Distribution Function of the Survival Times of Male Patients}
\label{fig:Weibull_Cumulative_distribution}
\end{figure}

Once we have the analytical form that characterizes the probabilistic behavior of the survival time of the male patients, the Cumulative Distribution Function (CDF) for male patients ($F_M(t)$) can be derived as follows in Equation~\ref{Weibul_Cumulative} and its graphical illustration for male patients is by Figure~\ref{fig:Weibull_Cumulative_distribution}.
\begin{equation*}
    F_M(t|\theta,\eta,\tau) = P(T \leq t) = \int_{0}^{t} f_M(x|\theta,\eta,\tau)\,dx\quad, \quad\quad t>0,
    \label{Cumulative_1}
\end{equation*}
\begin{equation}
    F_M(t|\theta,\eta,\tau) = 1 - e^{-(\frac{t-\tau} \theta)^\eta}.
    \label{Weibul_Cumulative}
\end{equation}

Identifying the CDF for a given random variable is a crucial achievement in probabilistic analysis because the CDF determines the probability or likelihood that a random variable will take on a specific value or less than that value. Thus, one can estimate the probability that an individual patient who underwent DBS for PD will survive until time $t$. Suppose a PD patient who underwent DBS wants to know what the probability is that he will live until 140 months. That is $P(t \leq 140) \approx 0.88$ as sketched in Figure~\ref{fig:Weibull_Cumulative_distribution}.

\begin{figure}[hbt!]
    \centering
    \includegraphics[width=5in]{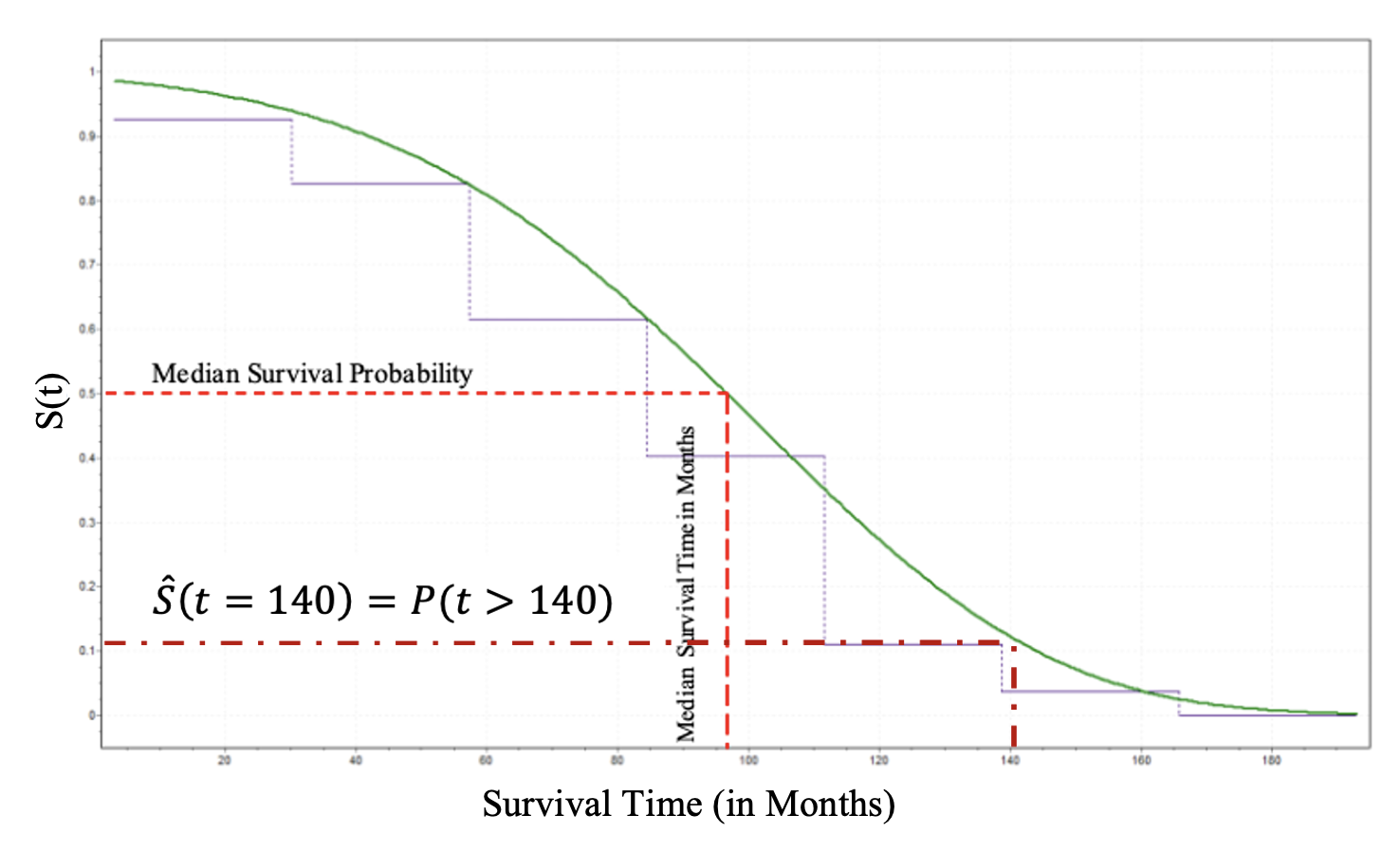}
    \caption{Survival Function of Male Patients.}
    \label{fig:Weibull_survival_distribution}
\end{figure}

Furthermore, the patient can estimate his/her survival probability beyond 140 months by utilizing the survival function. Since we have already estimated the CDF, the derivation of the Survival Function for the male patients and its analytical form is given by Equation~\ref{survival function for males} and its behavior by Figure~\ref{fig:Weibull_survival_distribution}. 
\begin{equation*}
    S_M(t|\theta,\eta,\tau) = 1- F_M(t|\theta,\eta,\tau) = P(T > t) = \int_{t}^{\infty} f_M(x|\theta,\eta,\tau)\,dx\quad, \quad\quad t>0,
    \label{survival_function}
\end{equation*}
\begin{equation}
    S_M(t|\theta,\eta,\tau) = e^{-(\frac{t-\tau} \theta)^\eta}.
    \label{survival function for males}
\end{equation}

Thus, as shown in Figure~\ref{fig:Weibull_survival_distribution} an individual patient surviving beyond 140 months can be estimated using Equation~\ref {survival function for males}, and that is approximately $0.12$. \\

According to the summary of the goodness of fit test results given in Table~\ref{Goodness_of_fit_Coefficient_ table}, the survival times of the Female patients are characterized by the \textbf{Lognormal (3P) distribution} and its analytical form is given by the following Equation~\ref {Lognormal_distribution_pdf}.

\begin{equation}
    f_F(x|\mu, \sigma, \gamma) = \frac{1}{(x-\gamma)\sigma \sqrt{2\pi}}exp\{-\frac{[ln(x-\gamma)-\mu]^2}{2\sigma^2}\}, \hspace{2em} \gamma < x,\, -\infty<\mu <\infty,\, \sigma > 0,
    \label{Lognormal_distribution_pdf}
\end{equation}

where $\mu$, $\gamma$, and $\sigma$ are Scale, Location, and Shape parameters of the 3P-Lognormal distribution, respectively. We have utilized the Maximum Likelihood Estimation (MLE) approach to estimate those three parameters of the Lognormal distribution as illustrated in the literature,~\cite{parameter_3p_log_normal_estimation},~\cite{3p_lognormal_parameter_est}. Thus, the estimated parameters are $\mu$ = 7.138, $\sigma$ = 0.03655, $\gamma$ = -1149.2. The graphical visualization of the shape of the Lognormal distribution is given by the following Figure~\ref{Lognormal_distribution_pdf}. For example, one can interpret the estimated probability that a PD patient will survive between 50 months and 100 months by using the probability distribution of the survival time of the female patients as shown in Figure~\ref{Lognormal_distribution_pdf}. That is, $P( 50 \leq t \leq 100) = 0.0085-0.0032= 0.0053$.

\begin{figure}[hbt!]
    \centering
    \includegraphics[width=5in]{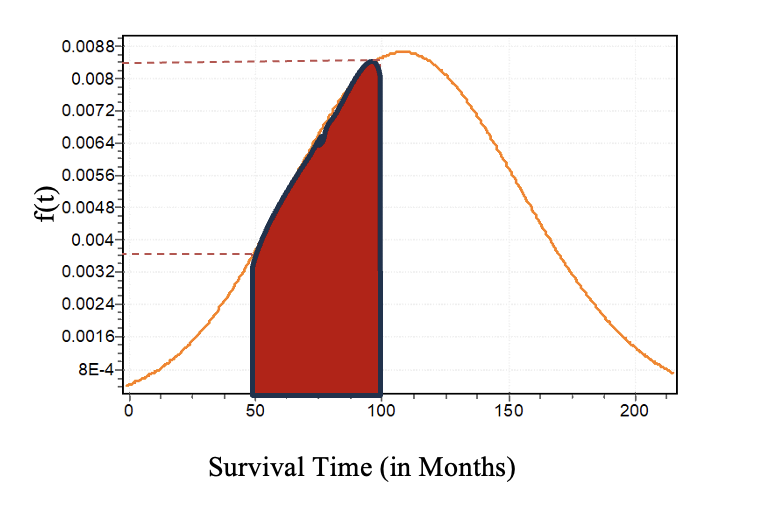}
    \caption{Probability Distribution of Survival Times of Female Patients.}
    \label{fig:Lognormal_pdf_distribution}
\end{figure}

\begin{figure}[hbt!]
    \centering
    \includegraphics[width=5in]{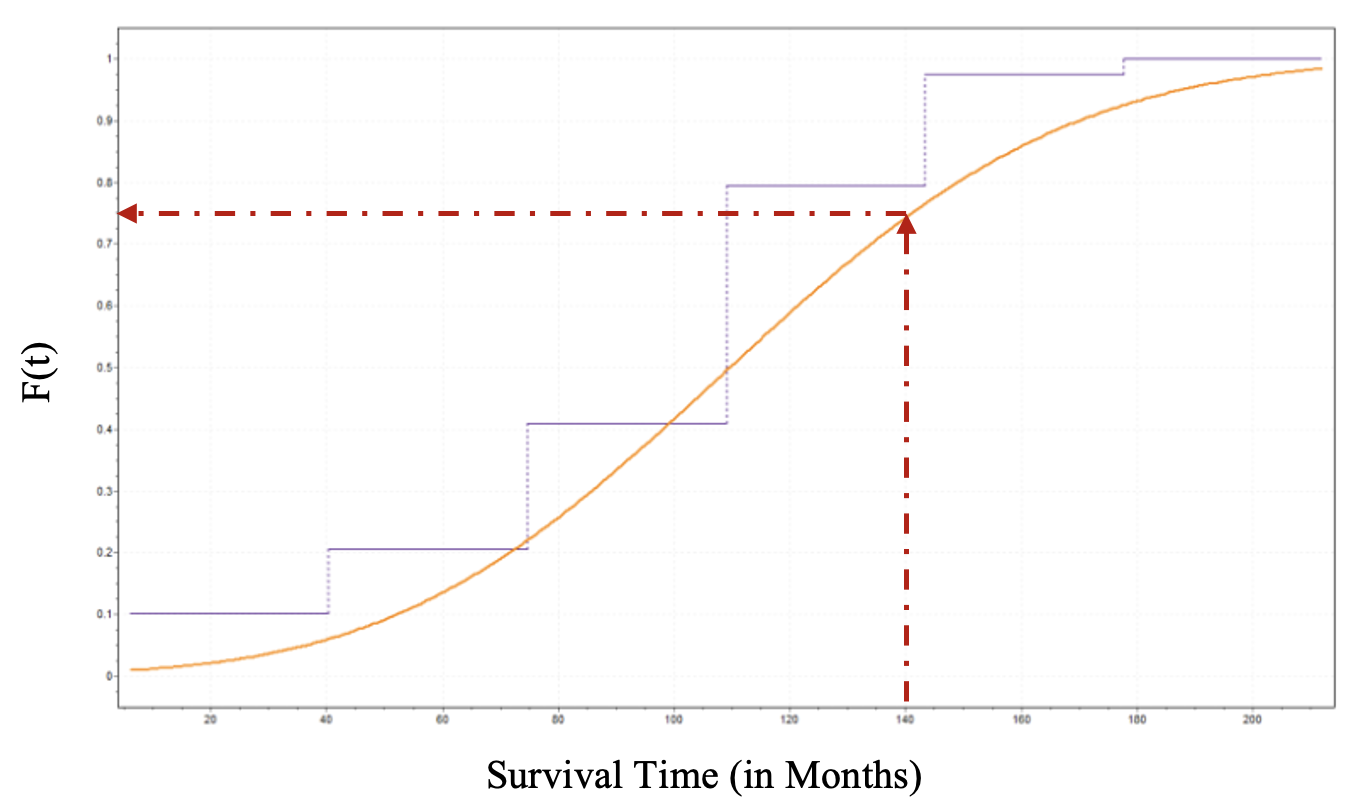}
    \caption{Cumulative Distribution Function of Survival Times of Female Patients}
    \label{fig:Lognormal_Cumulative_distribution}
\end{figure}

Once we derived the analytical form that characterizes the probabilistic behavior of the survival times of the female patients, the Cumulative Distribution Function (CDF) for female patients ($F_F(t)$) can be defined as the following Equation~\ref{Cumulative_DF_F}, given below, and its graphical illustration for female patients is given by Figure~\ref{fig:Lognormal_Cumulative_distribution}. We can use this CDF to estimate the probability of survival of the female patients until time $t$ $P(t\leq T)$. For example, suppose that we want to estimate the probability of surviving up to 140 months ($P( t \leq 140)$) using the CDF of the female patients given in Equation~\ref{Cumulative_DF_F}. Then the estimated probability is $P( t \leq 140) \approx 0.75$. The graphical visualization of the estimated value is shown in Figure~\ref{fig:Lognormal_Cumulative_distribution}.

\begin{equation}
    F_F(t|\mu, \sigma, \gamma) = \phi\left( \frac{ln(t-\gamma)-\mu}{\sigma} \right).
    \label{Cumulative_DF_F}
\end{equation}

\begin{figure}[hbt!]
    \centering
    \includegraphics[width=5in]{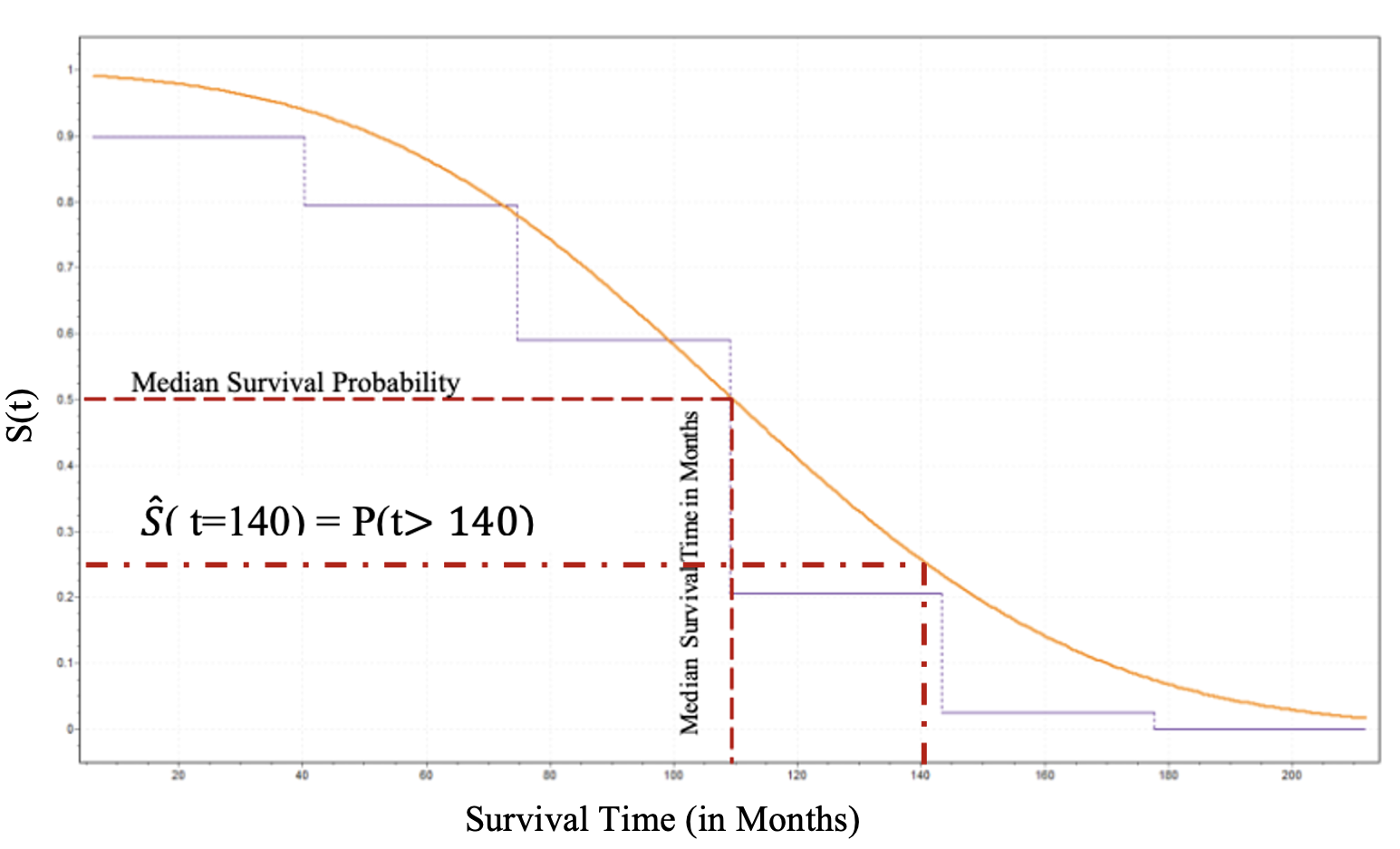}
    \caption{Survival Function for Female Patients.}
    \label{fig:lognormal_survival_distribution}
\end{figure}

\newpage
Finally, the estimated CDF can be used to derive the Survival Function for the female patients, and its analytical form is given by Equation~\ref{survival function for females} and its graphical visualization of the behavior by Figure~\ref {fig:lognormal_survival_distribution}.

\begin{equation}
    S_F(t|\mu, \sigma, \gamma) = 1- F_F(t|\mu, \sigma, \gamma) = 1- \phi\left( \frac{ln(t-\gamma)-\mu}{\sigma} \right).
    \label{survival function for females}
\end{equation}

Concerning the graphical illustration of the survival functions of the male and female patients, it is noticeable that the female patients have significantly higher median survival time than the median survival time of the male patients. This outcome is confirmed by the univariate results obtained from the KM survival analysis. However, the univariate survival analysis was performed under the assumption that all patients died from the same cause regardless of their other information. 
%\subsection{Multivariate Results}

\subsection{Multivariate Survival Analysis}
\label{Multivariate Survival Analysis}
\subsubsection{Cox-PH Model}

The Multivariate Cox-Ph model developing procedure was started with four individual risk factors that are identified by the medical professionals and believed to significantly contribute to the survival of a patient who underwent DBS. Thus, we initialized the model development procedure with four individual risk factors for both genders, but the performance of the models was not impressive. Thus, we utilized all individual risk factors and their two-way interactions along with the step-wise elimination method. Equation~\ref{Male_M} expresses the analytical form of the final cox-proportional hazard model for male patients, and Equation~\ref{Female_M} gives the final analytical form for the female patients.

\begin{equation}
    \widehat{h_M(t)} = h_0(t) exp(\hspace{0.3em}0.03074\hspace{0.3em} \textbf{X}_1 - 2.90540\hspace{0.3em} \textbf{X}_{2(1)} - 3.09802\hspace{0.3em} \textbf{$\sqrt{X_3}$} + 2.63631\hspace{0.3em} \textbf{X}_{2(1)}:\textbf{$\sqrt{X_3}$} ),
    \label{Male_M}
\end{equation}

\begin{equation}
\widehat{h_F(t)} = h_0(t) exp(\hspace{0.3em}0.17810\hspace{0.3em}\textbf{X}_1 + 2.45162\hspace{0.3em}\textbf{X}_{2(1)} -2.07761\hspace{0.3em}\textbf{X}_3),
\label{Female_M}
\end{equation}

\vspace{1em}
where \textit{$h_M(t)$} is the hazard function for male patients at time \textit{t},

\hspace{3em}\textit{$h_F(t)$} is the hazard function for female patients at time \textit{t},
 
\hspace{3em}\textit{$h_0(t)$} is the arbitrary baseline hazard function,
 
 \textit{$X_{i(l)}$} is the $i^{th}$ covariate with level \textit{l} (for a categorical risk factor) as described in Section 2.
 
  The above analytical models can be rearranged into a linear relationship between risk factors and the logarithm of the relative hazard ratios as follows for males and females, respectively.

  \begin{equation}
      log \left (\frac{\widehat{h_M(t)}}{h_0(t)} \right ) = \hspace{0.3em}0.03074\hspace{0.3em} \textbf{X}_1 - 2.90540\hspace{0.3em} \textbf{X}_{2(1)} - 3.09802\hspace{0.3em} \textbf{$\sqrt{X_3}$} + 2.63631\hspace{0.3em} \textbf{X}_{2(1)}:\textbf{$\sqrt{X_3}$}
  \end{equation}

  \begin{equation}
      log \left (\frac{\widehat{h_F(t)}}{h_0(t)} \right ) = \hspace{0.3em}0.17810\hspace{0.3em}\textbf{X}_1 + 2.45162\hspace{0.3em}\textbf{X}_{2(1)} -2.07761\hspace{0.3em}\textbf{X}_3
  \end{equation}
%Table with coefficients

The Table ~\ref{Coefficient_ table}, given below, consists of the information about the developed Cox-PH models for male patients and female patients, including the estimates of model coefficients (parameters), hazard ratios (exp(coef)), standard errors of coefficients, and 95\% lower and upper confidence intervals for the hazard ratios.
Also, Table ~\ref{hypothesis test}, given below, provides the global statistical significance values for the overall significance of the proposed Cox-PH models by using three hypothesis tests: Likelihood ratio test, Wald test, and Score test and all the p-values ($<< 0.05$) uniformly confirm that the proposed Cox-PH models are statistically significant at 5\% level of significance. These three tests are asymptotically equivalent. However, for small samples, the likelihood ratio test slightly outperforms the Wald test and both outperform the Score test.\\

\begin{table}[!h]
\begin{center}
\begin{tabular}{cccccc}
    \hline   
      Factor & Coefficient  & Hazard Ratio (exp(coef)) & SE(coef) & 95\% Lower & 95\% Upper  \\
     \hline
     \hline
    &   & \textbf{Male Patients}\\
     \hline
     $X_1$ & 0.050 & 1.051 & 0.015 & 1.022 & 1.082 \\
     
     $X_{2(1)}$ & -2.74  & 0.064 & 0.633 & 0.019 & 0.223\\
     
     $\sqrt{X_3}$ &  -3.23 & 0.039 & 0.517 & 0.014 & 0.109\\

     $X_{2(1)}:\sqrt{X_3}$ & 2.65 & 14.129 & 0.524 & 5.056 & 39.488 \\
     \hline
     &   & \textbf{Female Patients}\\
     \hline
     $X_1$ & 0.114 & 1.12 & 0.032 & 1.05 & 1.193\\
          % we can use this when the table elements are too long
     
     $X_{2(1)}$ & 1.014 & 2.76 & 0.496 & 1.04 & 7.293 \\
     
     $X_3$ & -0.474 & 0.6224 & 0.172 & 0.445 & 0.871 \\
     \hline
\end{tabular}
\end{center}
\caption{Coefficients and Hazard Ratios of the Proposed Cox-PH Models for Male Patients and Female Patients.}
\label{Coefficient_ table}
\end{table}

\begin{table}[h!]
    \centering
    \begin{tabular}{cccc}
        \hline
         & \hspace{6cm}\textbf{P-value} &  \\
       \textbf{Hypothesis Test} & \textbf{Male Patients} & \textbf{Male Patients} \\
        \hline
        \hline
        Likelihood ratio test & 3e-11 &  1e-4 \\
        Wald test & 6e-10 & 0.001  \\
        Score (log-rank) test & 2e-8 & 6e-4 \\
        \hline
    \end{tabular}
    \caption{\label{tab:hypothesis} Global Statistical Significance of the Proposed Cox-PH Models.}
    \label{hypothesis test}
\end{table}

Since the hazard ratio (exp(coef)) of the Cox-PH model measures the relative risk, it is possible to interpret the model coefficients in a meaningful way. If the model coefficient is positive ($\hat{\beta} > 0$), i.e., hazard ratio, (exp(coef)) $>$ 1, then the event of occurrence is high with a bad prognostic factor. However, if the coefficient is negative ($\hat{\beta} <$ 0), i.e., hazard ratio, (exp(coef)) $<$ 1, then the event of occurrence is low with a good prognostic factor. In the case of proposed models, both male and female patients consist of the coefficient of $X_1$ ($\hat{\beta}_{MX1} = 0.050 > 0$ and $\hat{\beta}_{FX1} = 0.114 > 0$) and are positive. It means that increasing age at implant of both male and female patients has a higher risk for death with a bad prognosis. In the case of the coefficients of $X_2$, the behavior of the two models is different from each other. For the female patients $X_2$ ($\hat{\beta}_{FX2} = 1.014 > 0$) has positive coefficient and has higher risk for death with a bad prognostic while male patients $X_2$ ($\hat{\beta}_{MX2} = -2.74 < 0$) has negative coefficient and the event of occurrence is low risk with a good prognostic factor.

\subsubsection{Validation of the Assumptions}
\label{Validation of the Assumptions}

In this section, we assess the five major assumptions of a Cox-PH model that need to be satisfied to perform a high-quality and robust survival analysis.

\begin{enumerate}

\item \textbf{Proportional hazards assumption}

We have performed the statistical test and also the graphical visualization based on Scaled Schoenfeld Residuals to validate the proportional hazards assumption on the Cox-PH model. Table~\ref{Proportiona hazard assumption}, given below, consists of p-values for the test based on scaled Schoenfeld residuals. The P-values for all covariates are greater than $0.05$ at the 5\% level of significance, confirming the fact that our proposed models for male and female patients satisfy the proportional hazards assumption. 

\begin{table}[!h]
\begin{center}
\begin{tabular}{cc}
    \hline   
      Covariate & P-value    \\
     \hline
     \hline
       & \textbf{Male Patients}\\
     \hline
     $X_1$ & 0.83 \\
     
     $X_{2(1)}$ & 0.78 \\
     
     $\sqrt{X_3}$ &  0.11\\

     $X_{2(1)}:\sqrt{X_3}$ & 0.24\\
     \hline
       & \textbf{Female Patients}\\
     \hline
     $X_1$ & 0.26 \\
          
     $X_{2(1)}$ & 0.68 \\
     
     $X_3$ & 0.51\\
     \hline
\end{tabular}
\end{center}
\caption{P-values for Testing the Proportional Hazards Assumption for Male and Female patients.}
\label{Proportiona hazard assumption}
\end{table}

Next, we visualize the graphical approaches of the scaled Schoenfeld residuals given in Figure~\ref{fig:schoenfeld_male} and Figure~\ref{fig:schoenfeld_female} for Male patients and Female patients, respectively. Both figures indicate a zero slope pattern by confirming the proportional hazard assumption.

\begin{figure}[hbt!]
\centering
\includegraphics[width=5.5in]{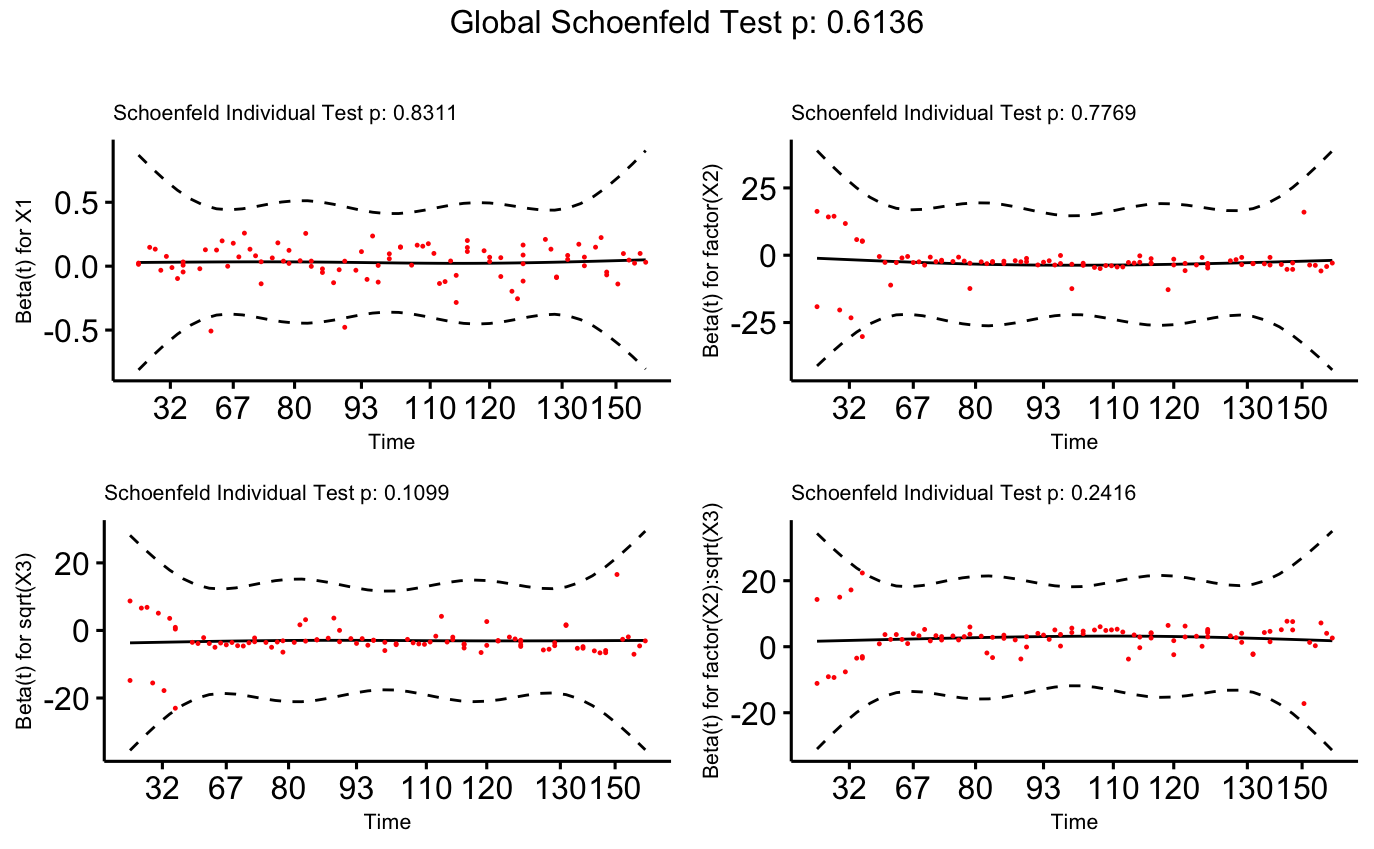}
\caption{Plots of Scaled Schoenfeld Residuals for Each Term of The Proposed Cox-PH Model for Male Patients}
\label{fig:schoenfeld_male}
\end{figure}

\begin{figure}[hbt!]
\centering
\includegraphics[width=5.5in]{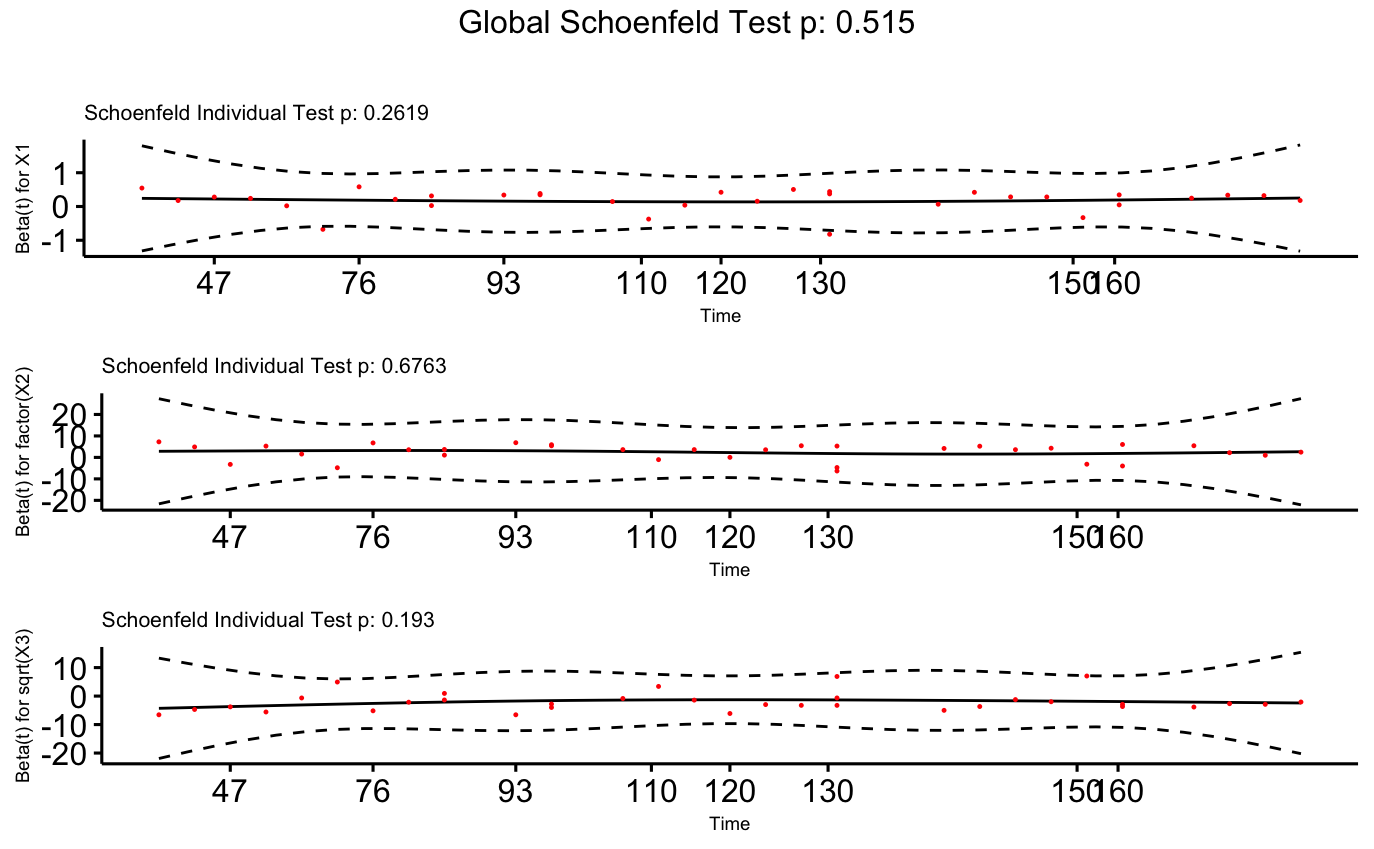}
\caption{Plots of Scaled Schoenfeld Residuals for Each Term of The Proposed Cox-PH Model for Female Patients.}
\label{fig:schoenfeld_female}
\end{figure}

\newpage
\item \textbf{Linear functional form of Continuous variables}

The linearity Assumption is one of the major underlying assumptions of the Cox-PH model that should be satisfied by the final model to achieve more robust results without misinterpretation. The graphical illustration of 
 the Martingale residuals for the continuous covariates can be used to identify a violation of this assumption. Figure~\ref{fig:linear_M} and Figure~\ref{fig:linear_F} uniformly confirm that both Cox-PH models given by Equations~\ref{Male_M} and ~\ref{Female_M} have satisfied the linearity assumption as given below for the continuous risk factors. 
\begin{figure}[hbt!]
    \centering
    \includegraphics[width=4in]{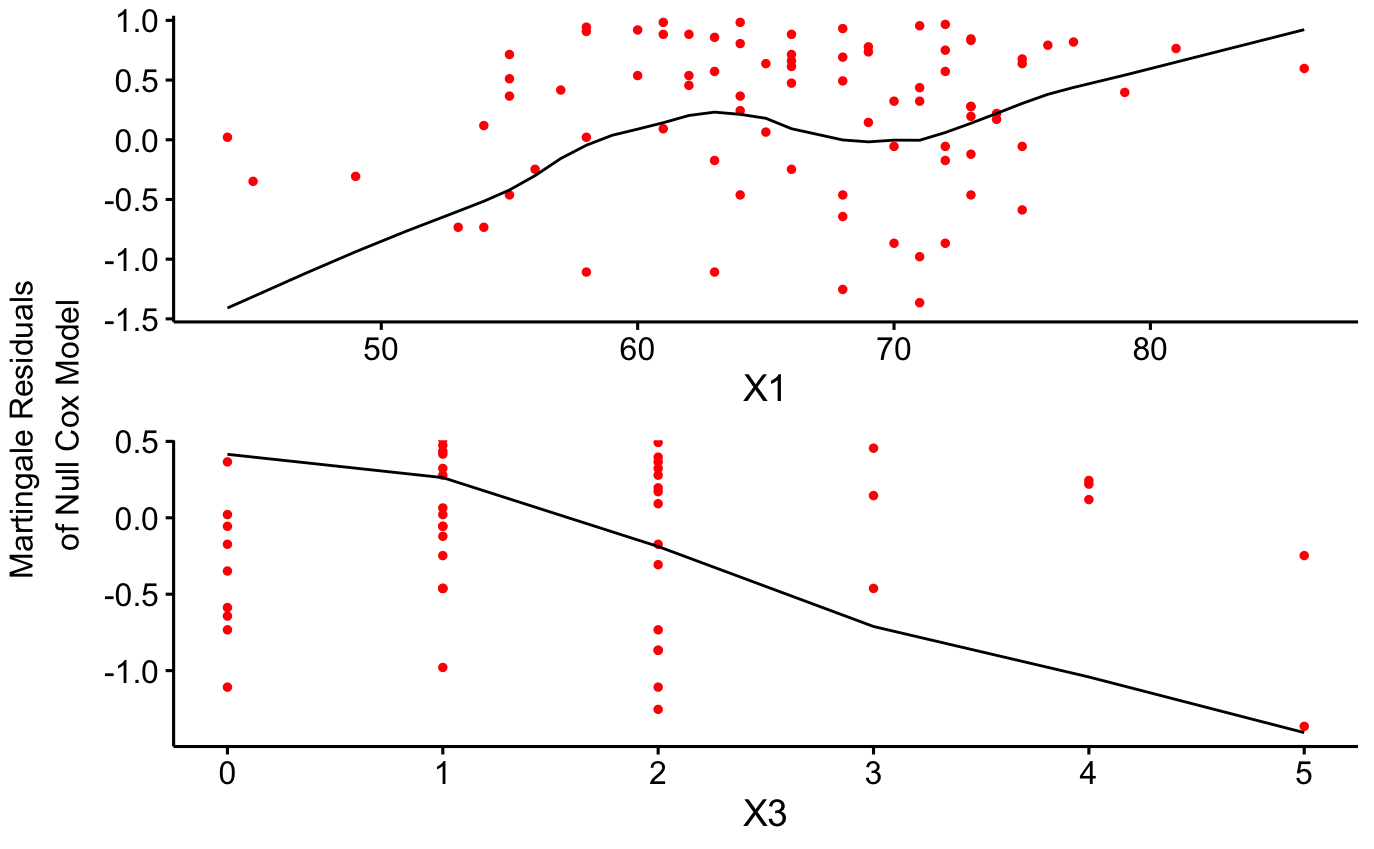}
    \caption{Martingale Residuals Plot for the Continuous Variables of Male Patients}
    \label{fig:linear_M}
\end{figure}

\begin{figure}[hbt!]
    \centering
    \includegraphics[width=4in]{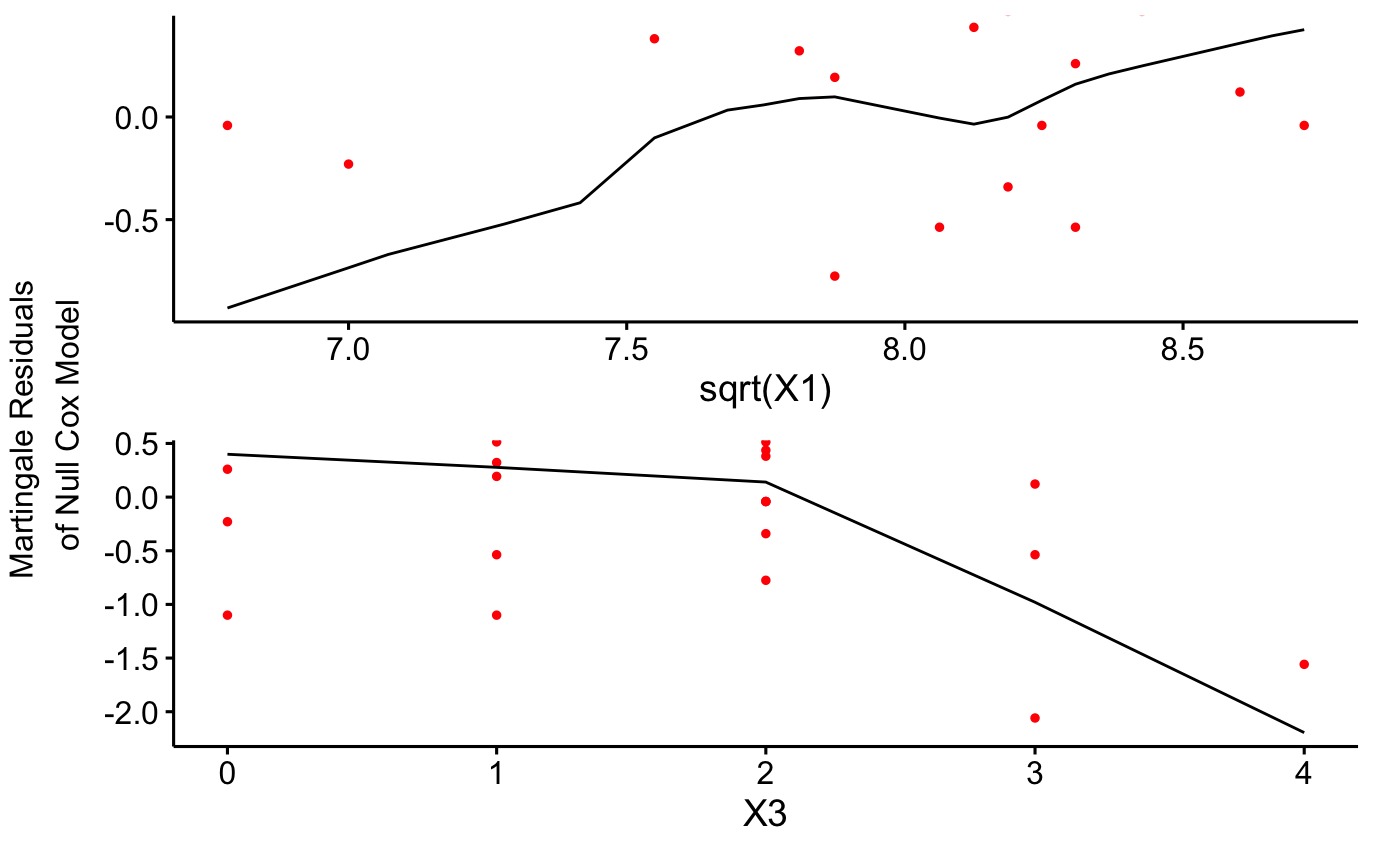}
    \caption{Martingale Residuals Plot for the Continuous Variables of Female Patients}
    \label{fig:linear_F}
\end{figure}

\newpage

\item \textbf{Check for Extreme Observations}

Under the next assumption, we check for the extreme values in our final Cox-PH model by utilizing deviance residual plots as given by Figure~\ref{fig:Influence_M} and ~\ref{fig:Influence_F}. Neither figures consist of any extreme outline observations. Thus, we can conclude that the proposed models have met the underlying assumption of no outliers or extreme observations.

\begin{figure}[hbt!]
    \centering
    \includegraphics[width=4.5in]{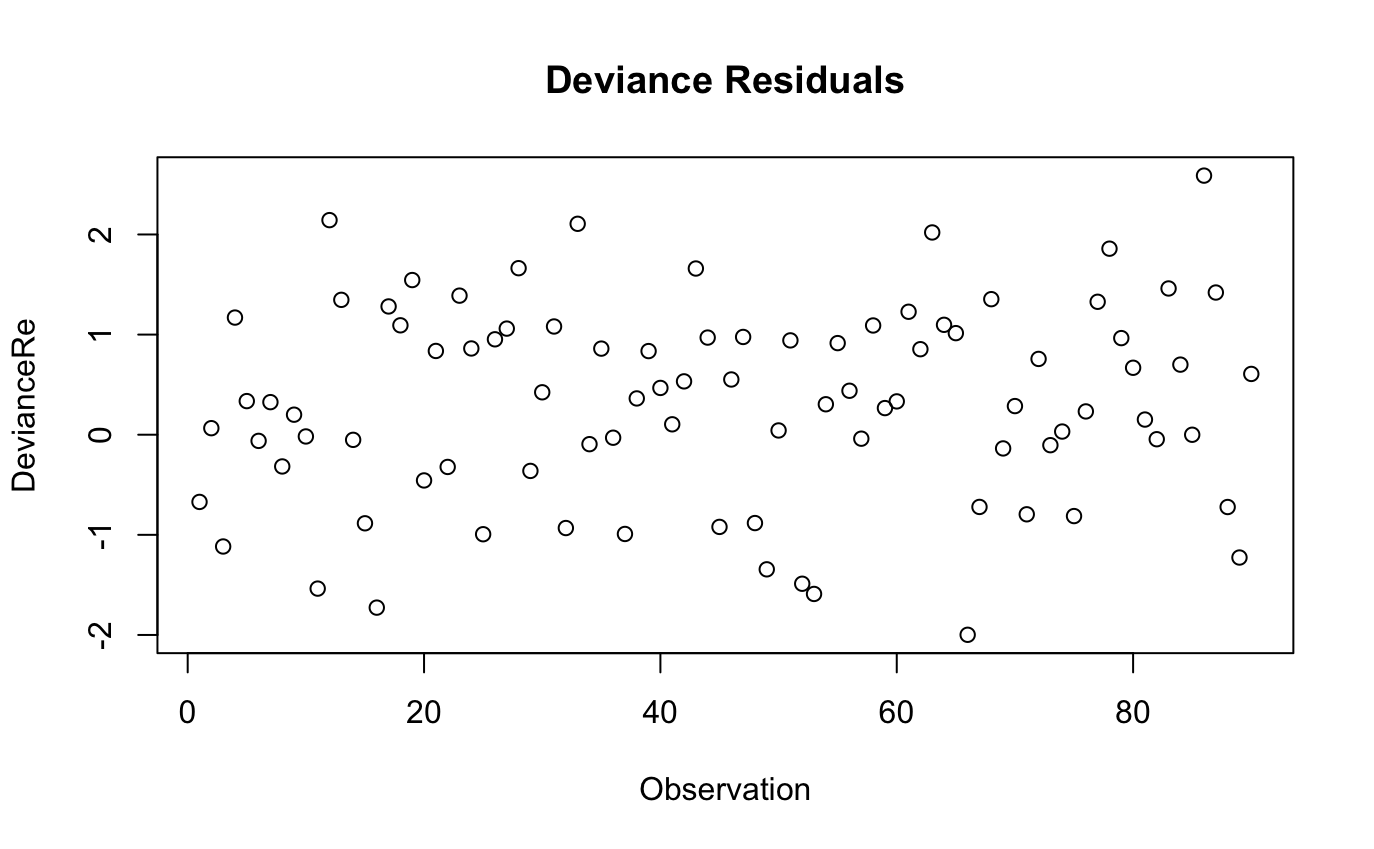}
    \caption{Deviance Residuals Plot for the Proposed Cox-PH Model for Male Patients}
    \label{fig:Influence_M}
\end{figure}

\begin{figure}[hbt!]
    \centering
    \includegraphics[width=4.5in]{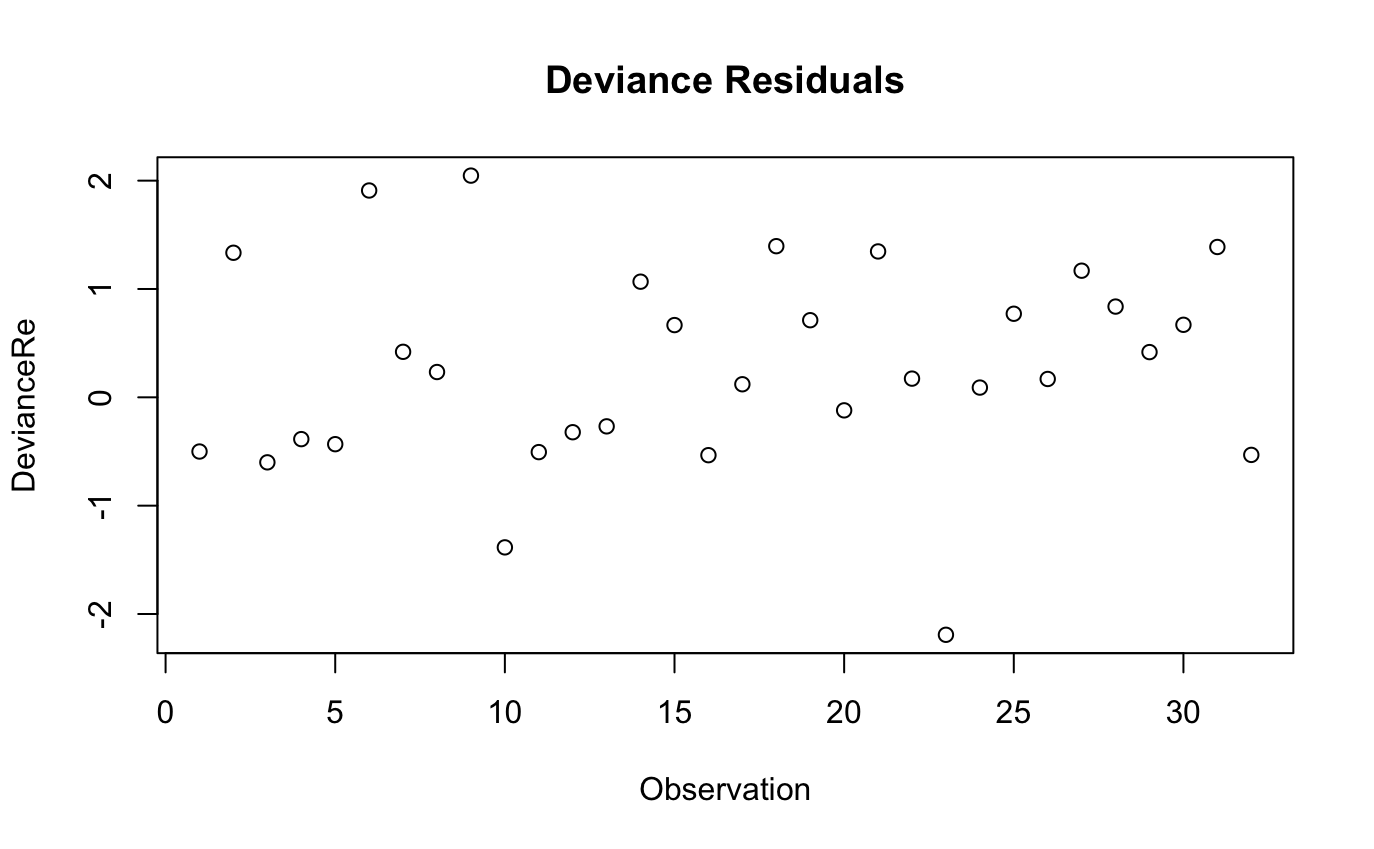}
    \caption{Deviance Residuals Plot for the Proposed Cox-PH Model for Female Patients}
    \label{fig:Influence_F}
\end{figure}

\item \textbf{Overall Fit of the Cox-PH Model}

Finally, we assessed the overall fit of the proposed Cox-PH models using the cumulative hazard of the Cox-Snell residuals plot. Figure~\ref{fig:CoxSnell_male} and Figure~\ref{fig:CoxSnell_female}, given below, graphically visualize the Cox-Snell residuals plots for both genders, and they follow a straight line with a zero intercept and a unit slope by confirming the high quality of the proposed models.

\begin{figure}[hbt!]
    \centering
    \includegraphics[width=4in]{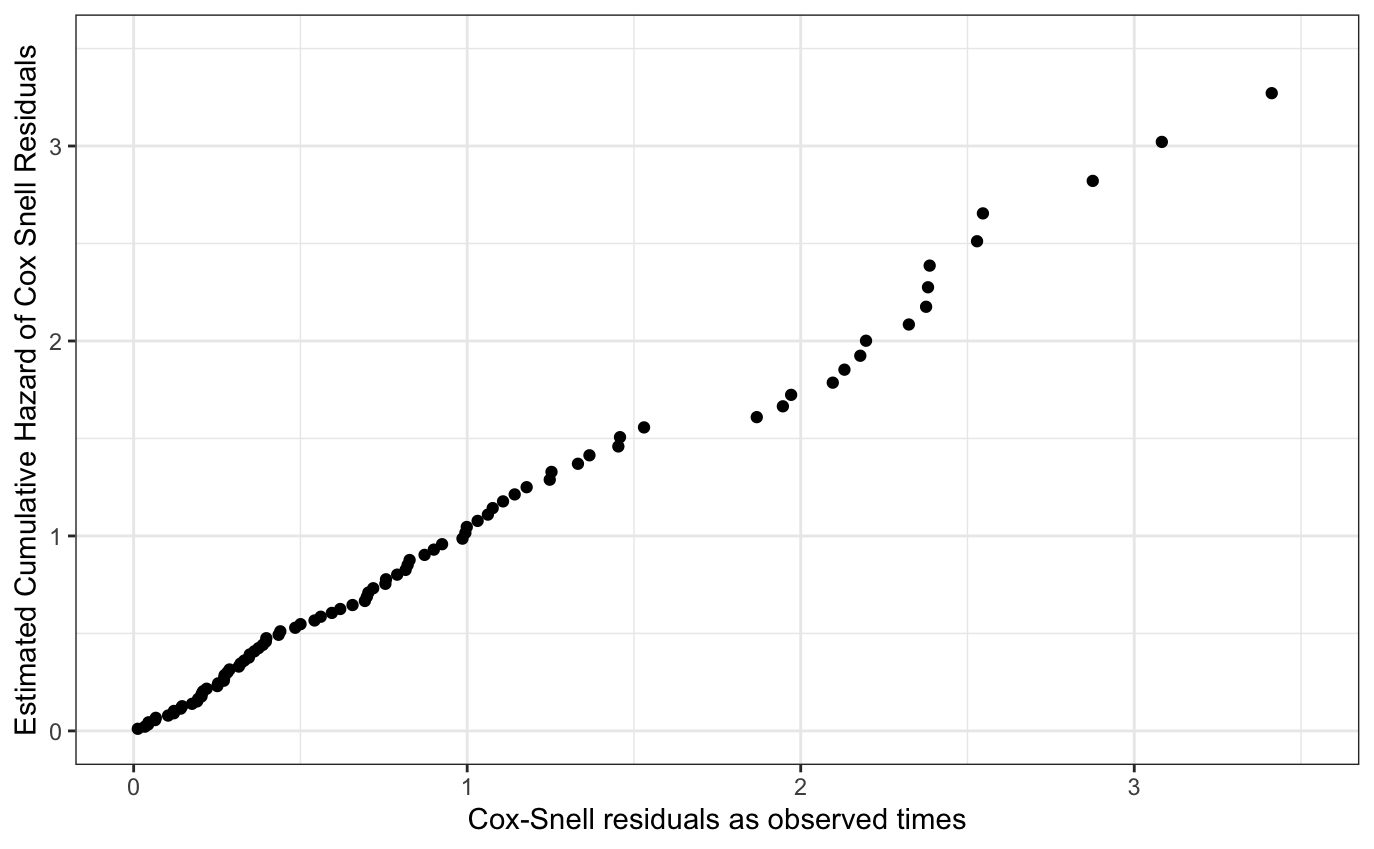}
    \caption{Cox-Snell Residuals Plot for the Proposed Cox-PH Model of Male Patients}
    \label{fig:CoxSnell_male}
\end{figure}

\begin{figure}[hbt!]
    \centering
    \includegraphics[width=4in]{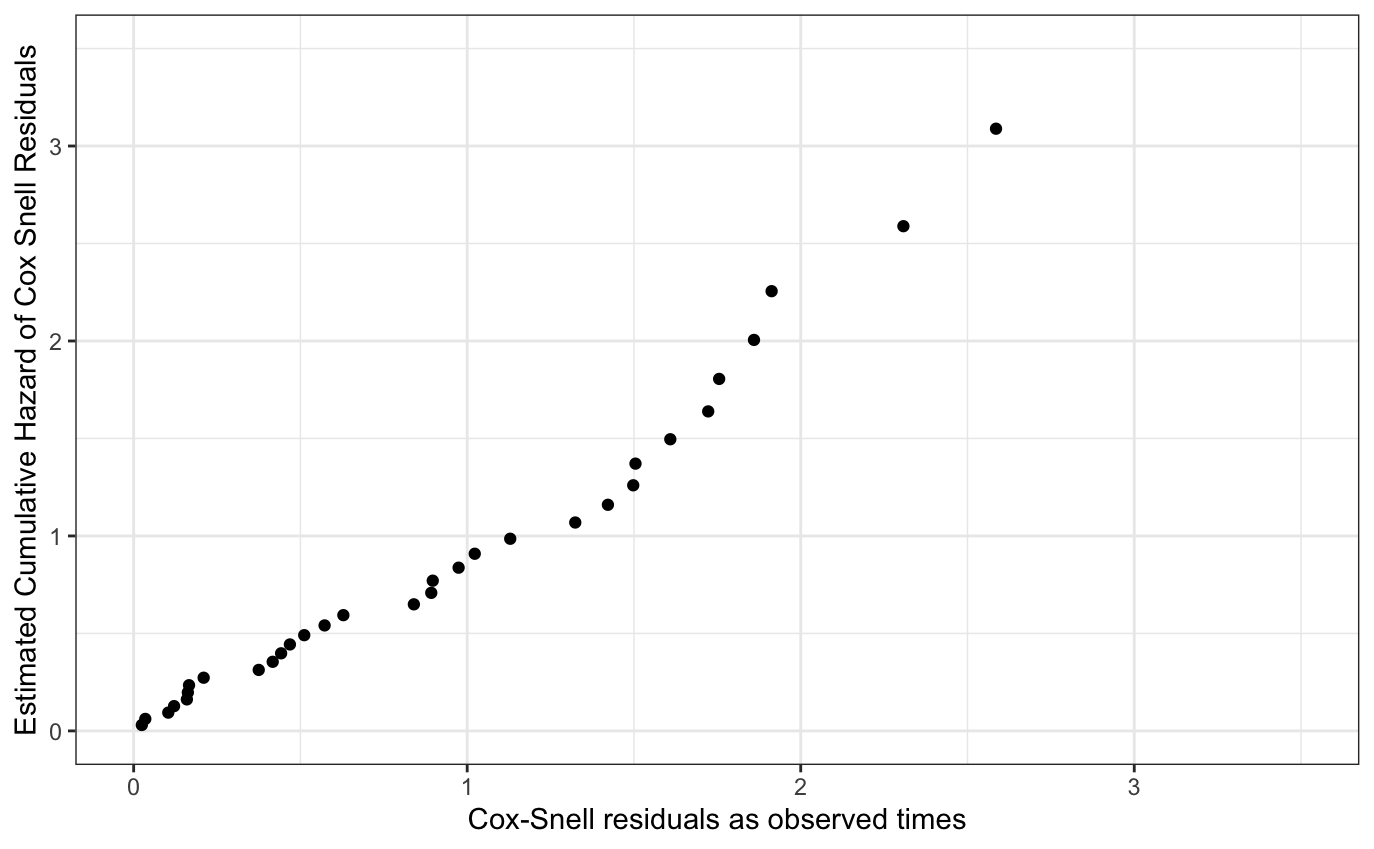}
    \caption{Cox-Snell Residuals Plot for the Proposed Cox-PH Model of Female Patients}
    \label{fig:CoxSnell_female}
\end{figure}

%\begin{figure}[hbt!]
 %   \centering
%    \includegraphics[width=4.5in]{Figures/influence_male.png}
 %   \caption{Deviance Residuals Plot for the Proposed Cox-PH Model for Male Patients.}
 %   \label{fig:Influence_M}
%\end{figure}

%\begin{figure}[hbt!]
 %   \centering
  %  \includegraphics[width=4.5in]{Figures/Influence_female.png}
   % \caption{Deviance Residuals Plot for the Proposed Cox-PH Model for Female Patients.}
   % \label{fig:Influence_F}
%\end{figure}

Thus, the proposed models for Male patients and Female patients have satisfied the underlying assumptions of the Cox-PH model and its high quality.
\end{enumerate}

\newpage
\section{Discussion and Conclusion}

In the scope of this paper, we have utilized nonparametric, semi-parametric, and parametric survival analysis to perform systematic and deep univariate and multivariate survival analysis for the patients who underwent DBS for PD. Within the first part of this study, the univariate survival analysis was performed through widely used nonparametric KM estimation method and more powerful and robust parametric approaches under the assumption that all patients deceased due to the same cause of death. Thus, we were able to compute and compare the survival probabilities using nonparametric KM and parametric methods for an individual patient. Since the survival times of the male and female patients indicated a statistically significant difference, as shown in Section~\ref{Results}, we performed survival analysis for male and female patients independently. In the sense of parametric survival analysis, 3-Parameter Lognormal Distribution probabilistically characterizes the survival times of the Female patients, while the probabilistic behavior of the survival times of the Male patients is characterized by 3-Parameter Weibull Distribution. 

Figures \ref{fig:comparison_survival_female} and \ref{fig:comparison_survival_male}, given below, graphically illustrate the comparison of the parametric and nonparametric survival curves for female patients and male patients, respectively. According to the graphical illustration, it is clear that the KM method underestimates the survival probabilities of the female patients while it overestimates the survival probabilities for the male patients. At this point in our study, we rely on survival probabilities estimated by the robust and more powerful parametric approach because of its robustness, as explained previously, compared to the widely used KM method. As shown in Figure~\ref{fig:comparison_survival_female}, the median survival time of the female patients is approximately 110 months, while in Figure \ref{fig:comparison_survival_male} the median survival time for male patients is approximately 95 months. Thus, we can conclude that the female patients who underwent DBS for PD have higher survival compared to that of male patients. However, in the case of univariate survival results, they are based on the assumption that all the patients' underlying cause of death is the same, without considering personalized risk factors or other relevant information.
\begin{figure}[hbt!]
    \centering
    \includegraphics[width=5in]{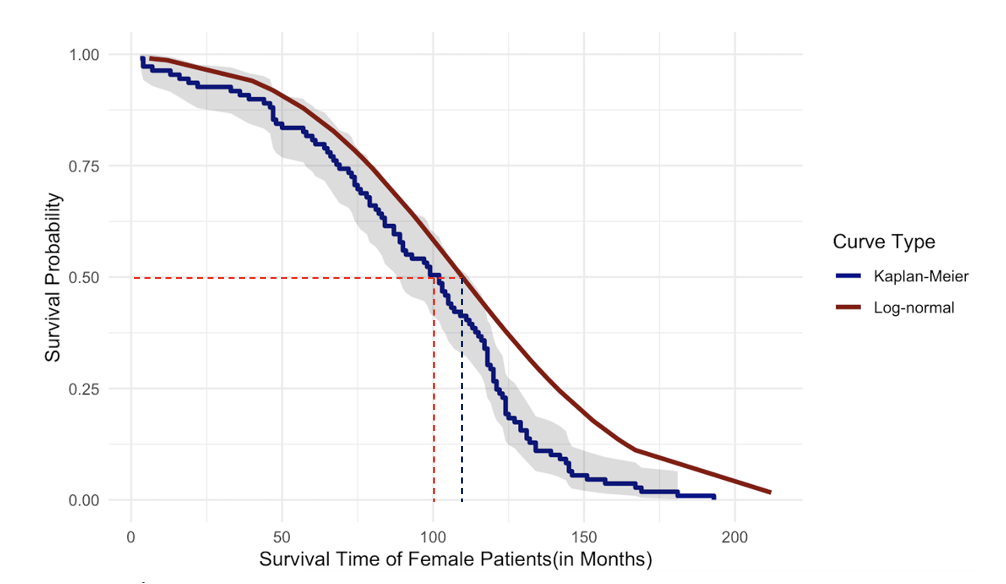}
    \caption{Comparison of Survival Probabilities for Female Patients Using Parametric and Nonparametric Approaches}
    \label{fig:comparison_survival_female}
\end{figure}

\begin{figure}[hbt!]
    \centering
    \includegraphics[width=5in]{Figures/Male_CPSN.png}
    \caption{Comparison of Survival Probabilities for Male Patients Using Parametric and Nonparametric Approaches}
    \label{fig:comparison_survival_male}
\end{figure}

Then, the effect of covariates was taken into account for both female and male patients separately, and we developed their own Cox-PH models. In the developing process, we considered the three individual covariates: Age at Implant, Initial Side of Implant, and Number of Revisions. Furthermore, the two-way interactions of the individual covariates are taken into account. However, it is important to note that in the literature, researchers do not consider interaction effects when developing a Cox-PH model. In Section \ref{Multivariate Survival Analysis}, we have developed the independent Cox-PH models for male patients and female patients. The final model for Male patients, given by Equation~\ref{Male_M}, consists of three individual statistically significant covariates: Age at Implant, Initial Side of Implant, the square root of the Number of revisions, and a statistically significant two-way interaction term between Initial Side of Implant and the square root of the Number of revisions. Moreover, the final Cox-PH model for female patients consists of only three individual statistically significant covariates: Age at Implant, Initial Side of Implant, and \textbf{Number of Revision} as given by Equation~\ref{Female_M}. In the model~\ref{Male_M}, the individual covariate, Initial side, has $HR < 1$. However, when it comes to the interaction between the Initial Side of Implant and the Number of Revision, it has $HR > 1$. This is a perfect example to illustrate the importance of adding interaction terms. Furthermore, in Section~\ref{Validation of the Assumptions}, we have addressed all underlying assumptions of the Cox-PH model, and they uniformly confirmed the robustness of our developed model and its high quality.\\

Once the Cox-PH model is finalized by addressing all underlying assumptions, the following important results are available concerning the long-term survival outcomes of the patients who underwent DBS for PD. 

\begin{itemize}
    \item For the covariate, Initial Side of Implant, the HR value (2.76) for female patients is greater than 1, which means the risk of death is higher with a bad prognosis. That is, Initial Right Side Implant has a higher risk of death than the Initial Left Side Implant. For the male patients, the HR value (0.064) is less than 1, which means the risk of death is lower with a good prognosis.
  \item For the interaction between Initial Side of Implant and the square root of Number of Revisions, the HR values are greater than 1, which means the risk of death is higher with a bad prognosis. That is, the number of Revisions that a patient has undergone has a stronger effect on the hazard ratio with the Right side of the Initial Implant compared to the Left side of the initial implant.

  \item Based on the hazard ratios, the most contributing individual covariates are Age at Implant, Initial Side of Implant, and Number of Revisions for male patients, and Initial Side of Implant, Age at Implant, and Number of Revisions for female patients, respectively. Furthermore, the ranking of the individuals and interaction terms for both models are given by Table~\ref{Ranking_according_to_hazard_ratio}.
\end{itemize}

\begin{table}[!h]
\begin{center}
\begin{tabular}{ccc}
    \hline   
      Covariate & Hazard Ratio(exp( Coeff) & Rank    \\
     \hline
     \hline
       & \textbf{Male Patients}\\
     \hline
     $X_{2(1)}:\sqrt{X_3}$ & 14.129  & 1\\
     
     $X_1$ & 1.051 & 2\\
     
     $X_{2(1)}$ &  0.064 & 3\\

     $\sqrt{X_3}$ & 0.039 & 4\\
     \hline
       & \textbf{Female Patients}\\
     \hline
     $X_{2(1)}$ & 2.76 & 1\\
          
     $X_1$ & 1.12 & 2\\
     
     $X_3$ & 0.6224 & 3\\
     \hline
\end{tabular}
\end{center}
\caption{Ranking of Contributing Covariates and Interactions Using Hazard Ratios}
\label{Ranking_according_to_hazard_ratio}
\end{table}
%\section{Reference}
\newpage
\section{Statements and Declarations}

\subsection*{Ethical approval:}
\begin{itemize}
    \item We, all authors, Dr. Malinda Iluppangama, Dilmi Abeywardana and Dr. Chris P. Tsokos have agreed on authorship, read and approved the manuscript, and given consent for submission and subsequent publication of the manuscript. Dr. Iluppangama acts as the corresponding author for this manuscript.
    \item We ensure that the manuscript and its contents are not a duplicate publication, and it is completely original.
    \item We ensure that the manuscript represents our original work and meets the ethical standards set by the Committee on Publication Ethics (COPE).
\end{itemize}

\subsection*{Consent to Publish declaration:}

The authors give full consent to BMC Neurology to publish this manuscript.

\subsection*{Consent to Participate declaration:} 

Not applicable

\subsection*{Ethics declaration:} 

Not applicable

\subsection*{Competing interests:}

The authors declare that they have no competing interests.

\subsection*{Data Availability:} 
The original contributions presented in this study are included in the article. Further inquiries can be directed to the corresponding author(s).

\subsection*{Author Contributions:} 

Conceptualization, Malinda Iluppangama and Chris Tsokos; methodology, Malinda Iluppangama and Dilmi Abeywardana; software, Malinda Iluppangama and Dilmi Abeywardana; validation, Malinda Iluppangama, Dilmi Abeywardana, and Chris Tsokos; formal analysis, Malinda Iluppangama and Dilmi Abeywardana; investigation, Malinda Iluppangama; resources, Chris Tsokos; writing original draft preparation, Malinda Iluppangama and Chris Tsokos; writing review and editing, Chris Tsokos; visualization, Malinda Iluppangama and Dilmi Abeywardana; supervision, Chris Tsokos; project administration, Chris Tsokos. All authors have read and agreed to the published version of the manuscript.

\subsection*{Conflicts of interest:}

The author(s) declared no potential conflicts of interest with respect to the research, authorship, and/or publication of this article.

%\newpage
\bibliographystyle{unsrt}
%\bibliography{url}
\bibliography{coxph.bib}
\end{document}